\documentclass[12pt]{oxarticle}
\pdfoutput=1

\usepackage{graphicx, float, array, xspace, amscd, amsmath, amsthm, amssymb, latexsym, relsize, bbm}
\usepackage[letterpaper, left=1.7in, right=0.3in, top=0.8in, bottom=1.2in]{geometry}
\usepackage[bbgreekl]{mathbbol}
\usepackage{amsfonts, tikz-cd}
\usepackage[T1]{fontenc}
\PassOptionsToPackage{linecolor=blue,backgroundcolor=blue!25,bordercolor=blue,textsize=scriptsize}{todonotes}
\usepackage{todonotes}
\usepackage{tikz}
\usetikzlibrary{backgrounds}
\usepackage[framemethod=tikz]{mdframed}
\AtBeginEnvironment{mdframed}{%
\tikzset{every picture/.style={}}%
}
\mdfsetup{roundcorner=.5ex}
{\begin{mdframed}[backgroundcolor=white]}%
{\end{mdframed}}

\DeclareSymbolFontAlphabet{\mathbb}{AMSb}
\DeclareSymbolFontAlphabet{\mathbbl}{bbold}

\usepackage[sort&compress, comma, square, numbers]{natbib}
\usepackage[all,cmtip]{xy}
\usepackage{color}
\definecolor{MyDarkBlue}{rgb}{0.15,0.25,0.45}
\usepackage[linktocpage=true]{hyperref}
\hypersetup{colorlinks=true, citecolor=blue, linkcolor=blue, urlcolor=blue, pdfauthor={}, pdftitle={},breaklinks=true}

\let\SS=\S 

\renewcommand{\#}{^{\sharp}}

\newcommand{\Ric}{{\rm Ric}}

\newcommand{\LC}{\text{\tiny LC}}

\renewcommand{\sb}{{\overline{\sigma}}}

\newcommand{\rb}{{\overline{ r}}}

\newcommand{\Ob}{{\overline{ \Omega}}}

\newcommand{\w}{{\,\wedge\,}}
\newcommand{\wt}{\widetilde}
\newcommand{\wh}{\widehat}

\newcommand{\fD}{{\mathfrak{D}}}

\newcommand{\half}{\frac{1}{2}}

\newcommand{\ab}{{\overline\alpha}}
\newcommand{\bb}{{\overline\beta}}

\newcommand{\db}{{\overline\delta}}

\renewcommand{\a}{\alpha}
\renewcommand{\b}{\beta}
\newcommand{\G}{\Gamma}
\renewcommand{\d}{\delta}\newcommand{\D}{\Delta}
\newcommand{\ve}{\varepsilon}

\renewcommand{\th}{\theta}\newcommand{\Th}{\Theta}

\renewcommand{\k}{\kappa}
\renewcommand{\l}{\lambda}
\newcommand{\m}{\mu}
\newcommand{\n}{\nu}
\newcommand{\x}{\xi}

\renewcommand{\r}{\rho}
\newcommand{\s}{\sigma}\renewcommand{\S}{\Sigma}
\renewcommand{\t}{\tau}

\newcommand{\vph}{\varphi}

\renewcommand{\o}{\omega}\renewcommand{\O}{\Omega}

\DeclareFontFamily{OT1}{pzc}{}
\DeclareFontShape{OT1}{pzc}{m}{it}{<-> s * [1.200] pzcmi7t}{}
\DeclareMathAlphabet{\mathpzc}{OT1}{pzc}{m}{it}

\newcommand{\cA}{\mathcal{A}}
\newcommand{\ccB}{\mathpzc B}

\newcommand{\cD}{\mathcal{D}}\newcommand{\ccD}{\mathpzc D}
\newcommand{\ccE}{\mathpzc E}
\newcommand{\cF}{\mathcal{F}}

\newcommand{\cH}{\mathcal{H}}\newcommand{\ccH}{\mathpzc H}

\newcommand{\cM}{\mathcal{M}}\newcommand{\ccM}{\mathpzc M}

\newcommand{\cO}{\mathcal{O}}

\newcommand{\cQ}{\mathcal{Q}}
\newcommand{\cR}{\mathcal{R}}

\newcommand{\ccT}{\mathpzc T}

\newcommand{\ccX}{\mathpzc X}
\newcommand{\ccY}{\mathpzc Y}
\newcommand{\cZ}{\mathcal{Z}}\newcommand{\ccZ}{\mathpzc Z}

\newcommand{\ccZb}{{\overline \ccZ}}
\newcommand{\ccZt}{{\wt \ccZ}}
\newcommand{\ccZtb}{{\overline{{\wt \ccZ}}}}

\DeclareFontFamily{U}{bbold}{}
\DeclareFontShape{U}{bbold}{m}{n}
 {  <-5.5> s*[1.05] bbold5
    <5.5-6.5> s*[1.05] bbold6
    <6.5-7.5> s*[1.05] bbold7
    <7.5-8.5> s*[1.05] bbold8
    <8.5-9.5> s*[1.05] bbold9
    <9.5-11.5> s*[1.05] bbold10
    <11.5-16> s*[1.05] bbold12
    <16-> s*[1.05] bbold17
 }{}

\newcommand{\IR}{\mathbbl{R}}

\font\eightrm=cmr8 at 8pt
\font\csc=cmcsc10

\newcommand{\beq}{\begin{equation}}
\newcommand{\eeq}{\end{equation}}
\newcommand{\beqnn}{\begin{equation*}}
\newcommand{\eeqnn}{\end{equation*}}
\newcommand{\bea}{\begin{eqnarray}}
\newcommand{\eea}{\end{eqnarray}}
\newcommand{\bean}{\begin{eqnarray*}}
\newcommand{\eean}{\end{eqnarray*}}

\newcommand{\sref}[1]{\SS\ref{#1}}

\newcommand{\norm}[1]{\left\| #1\right\|}
\newcommand{\ee}{\text{e}}
\newcommand{\ii}{\text{i}}

\newcommand{\place}[3]{\vbox to0pt{\kern-\parskip\kern-7pt
                             \kern-#2truein\hbox{\kern#1truein #3}
                             \vss}\nointerlineskip}

\DeclareFontFamily{U}{wncy}{}
\DeclareFontShape{U}{wncy}{m}{n}{<->wncyr10}{}
\DeclareSymbolFont{mcy}{U}{wncy}{m}{n}
\DeclareMathSymbol{\sha}{\mathord}{mcy}{"58}

\newcommand{\del}{{\partial}}
\newcommand{\delb}{{\overline{\partial}}}

\newcommand{\lb}{{\overline\lambda}}
\newcommand{\nb}{{\overline\n}}
\newcommand{\mb}{{\overline\m}}

\newcommand{\Db}{{\overline D}}

\newcommand{\A}{\cA}

\newcommand{\mfb}{\mathfrak{b}}

\newcommand{\EndE}{{\text{End}\,E}}
\newcommand{\EndT}{{\text{End}\,\ccT_\ccX}}
\newcommand{\dd}{{\text{d}}}

\newcommand{\Hend}{H^1(\ccX,\text{End}\,E)}

\newcommand{\K}{K\"ahler\xspace}

\def\im{{\rm im ~}}

\def\ker{{\rm ker ~}}

\newcommand{\vol}{\dd^6 x \sqrt{g}\, }
\newcommand{\tr}{\text{Tr}\hskip2pt}

\newcommand{\tb}{{\overline{\tau}}}

\newcommand{\ap}{{\a^{\backprime}\,}}
\renewcommand{\sb}{{\overline{\sigma}}}

\renewcommand{\rb}{{\overline{\rho}}}
\renewcommand{\=}{\;=\;}

\newcommand{\cDb}{{\overline\cD}}
\makeatletter
\g@addto@macro\bfseries{\boldmath}
\makeatother

\newcommand{\citeM}{\cite{Candelas:2016usb}\xspace}
\newcommand{\citeSG}{\cite{McOrist:2019mxh}\xspace}
\newcommand{\citeMatter}{\cite{McOrist:2016cfl}\xspace}

\newcommand{\citeCOV}{\cite{Candelas:2018lib,Candelas:2016usb}\xspace}

\newcommand{\citeAQS}{\cite{Anguelova:2010ed}\xspace}
\newcommand{\citeOS}{\cite{delaOssa:2014cia}\xspace}

\newcommand{\citeS}{\cite{Ashmore:2018ybe}\xspace}

\renewcommand{\baselinestretch}{1.1}
\numberwithin{equation}{section}
\proofmodefalse
\begin{document}
\pagestyle{empty}      
\ifproofmode\underline{\underline{\Large Working notes. Not for circulation.}}\else{}\fi

\begin{center}
\null\vskip0.2in
{\Huge  Heterotic Quantum Cohomology\\[0.5in]}
{\csc   Jock McOrist$^{\dagger\,1}$ and Eirik Eik Svanes$^{*\,2}$\\[0.5in]}

{\it 
$^\dagger$Department of Mathematics\hphantom{$^2$}\\
School of Science and Technology\\
University of New England\\
Armidale, 2351, Australia\\[3ex]
$^*$ Department of Mathematics and Physics \\
Faculty of Science and Technology\\
University of Stavanger\\
N-4036, Stavanger, Norway\\
}
\footnotetext[1]{{\tt jmcorist@une.edu.au}}
\footnotetext[2]{{\tt eirik.e.svanes@uis.no}}
\vspace{1cm}
{\bf Abstract\\[-8pt]}
\end{center}

We reexamine the massless spectrum of a heterotic string vacuum at large radius and present two results. The first result is to construct a vector bundle $\cQ$ and operator $\cDb$ whose kernel amounts to deformations solving `F-term' type equations. This resolves a dilemma in previous works in which the spin connection is treated as an independent degree of freedom, something that is not the case in string theory. The second result is to utilise the moduli space metric, constructed in previous work, to define an adjoint operator $\cDb^\dag$. The kernel of $\cDb^\dag$ amounts to deformations solving `D-term' type equations. Put together, we show there is a vector bundle $\cQ$ with a metric, a $\cDb$ operator and a gauge fixing (holomorphic gauge) in which the massless spectrum are harmonic representatives of $\cDb$. This is remarkable as  previous work indicated the Hodge decomposition of massless deformations were complicated and in particular not harmonic except at the standard embedding.

\vskip150pt

\newgeometry{left=1.5in, right=0.5in, top=0.75in, bottom=0.8in}
%
\newpage
{\baselineskip=10pt\tableofcontents}
\restoregeometry
\setcounter{page}{1}
\pagestyle{plain}
\renewcommand{\baselinestretch}{1.3}
\null\vskip-10pt

\section{Introduction}
We study heterotic vacua that have large radius limits. In this limit, the vacua are solutions of $\ap$--corrected supergravity. This effective field theory is fixed up to and including $\ap^2$ by supersymmetry and the result is also consistent with string scattering amplitudes. We refer to the resulting equations of motion as the Hull--Strominger system \cite{Hull:1986kz, Strominger:1986uh}. The vacua we study have a smooth limit $\ap\to0$ which forces $H\to 0$ and a ten--dimensional spacetime  of the form $\IR^{3,1}\times \ccX$ with $\ccX$ a complex 3-fold with $c_1(\ccX)=0$. With an appropriate gauge fixing the dilaton is constant  up to order $\ap^3$ \cite{Anguelova:2010ed}  and the background is topologically a CY manifold with a holomorphic vector bundle $\ccE$ admitting a connection $A$ that satisfies the hermitian Yang--Mills equation.  These are the only compact supergravity solutions with a valid $\ap\to 0$ limit, which guarantees the supergravity solutions are also solutions of string theory up to non--perturbative corrections. There are solutions in which as $\ap\to0$, the three-form $H$ is non--trivial. However, one must study either non--compact manifolds, be non--perturbative in the dilaton via, e.g. non--geometric solutions, or consider solutions without a convergent $\ap$ expansion. We do not consider such solutions here because the higher order perturbative $\ap$ corrections to the Hull--Strominger equations (and likely worldsheet instantons) will play an important role modifying the quadratic order $\ap$ behaviour, and these $\ap$ corrections have not been completely determined yet. We expand upon our justification for restricting to solutions with a smooth $\ap\to0$ limit in Appendix \ref{app:NoGo}.

The moduli space is a finite dimensional complex \K manifold $\ccM$. Each point $y\in \ccM$ corresponds to a heterotic vacuum and there is a Kuranishi map which relates tangent vectors $\d y$ on $\ccM$ to deformations of fields. Small gauge transformations correspond to deforming the Kuranishi map and there is a choice in which the map is holomorphic \citeSG. We call this holomorphic gauge.   In holomorphic gauge, there is a relation between deformations of the connection on the tangent bundle $\ccT_\ccX$ and the  moduli of $\ccX$ \cite{Candelas:2018lib}
\beq\label{eq:spinmoduli}
 \d \Th^{(0,1)}  \= \d y^\a \fD_\a \Th~, \quad \text{where} \quad \fD_{\a}\Th{}_\mb{}^\n{}_{\s} \= \nabla_\s \, \D_{\a\mb}{}^\n + \ii \, \nabla^{\n} \, (\del_\a\o)_{\s\mb}  ~.
\eeq
 Here we have introduced  coordinates on $\ccM$ as $y^a = (y^\a, y^\bb)$ using the  complex structure inherited from the heterotic theory. On $\ccX$ real coordinates are denoted  $x^m$, complex coordinates $(x^\m, x^\nb)$ and $\nabla_\s$ is the usual Levi--Civita connection on $\ccX$. We denote $J$  the complex structure on $\ccX$ and (in holomorphic gauge) a holomorphic deformation is a $(0,1)$--form valued in the holomorphic tangent bundle. We write this in the notation of \citeM viz. $\d y^\a \D_{\a\,\mb}{}^\n$. Similarly, if $\o$ is the hermitian form on $\ccX$ then $(\del_\a \o)_{\s\mb}$ is a holomorphic deformation of the hermitian form restricted to its $(1,1)$ component. In holomorphic gauge this is equal (up to a factor of $\ii$) to a gauge invariant deformation of the B-field, denoted $\ccB_\a^{(1,1)}$ \citeSG. The analogue of complexified \K deformations for heterotic theories is the combination
 $$
 \ccZ_\a \= \ccB_\a + \ii \del_\a \o~.
 $$
  The key point is that $\d \Th^{(0,1)}$ is fixed in terms of $\D_\a$ and $\ccZ_\a^{(1,1)}$. In contrast, in  \cite{delaOssa:2014cia, Anderson:2014xha, Garcia-Fernandez:2015hja} the deformation $\d \Th^{(0,1)}$ is treated a degree of freedom independent of the other moduli with its own set of parameters. That is, $\d \Th^{(0,1)}$ is an arbitrary element of $H^1(\ccX,  \EndT)$ and with these spurious degrees of freedom one finds a  $\Db$--operator on a double extension bundle $Q$ in which $\Db^2 = 0$ and its cohomology is related to the deformations of heterotic theories plus the spurious degrees of freedom. A natural question  is to ask  what happens to the operator $\Db$ and bundle $Q$ when we eliminate the spurious degrees of freedom? Our initial goal is to answer this question. What we find is that in holomorphic gauge \citeSG there is an extension bundle $\cQ$ without the spurious degrees of freedom with a $\cDb$ operator acting on sections of this bundle. Its kernel amounts to `F-term' type equations. Our next goal is to use the moduli space metric constructed in \citeCOV\footnote{This is the natural string theory moduli space metric constructed in $\ap$--corrected supergravity and so is valid for the solutions of Hull--Strominger with a smooth $\ap\rightarrow0$ limit.} to construct an adjoint operator $\cDb^\dag$ and show that the co-kernel of $\cDb$ describe D-term type equations. D-term equations here refer to first order deformations of the Hermitian--Yang--Mills equation and balanced equations.\footnote{We use F-term equations and D-terms equations loosely: without an off-shell definition of string theory, viz. string field theory, they don't really make sense.}

 In other words, the physical massless moduli  of the heterotic supergravity at large radius are harmonic representatives of the cohomology of the $\cDb$--operator. 
 
 In the next section we review some results setting up our notation. In \sref{s:atiyah} we review how extension bundles describe deformations of complex manifolds. In \sref{s:spur} we revist and refine the calculation in \citeOS. We find a family of operator $\Db$ on a double extension, and using the metric in \citeM show its adjoint describes deformations of the Hermitian--Yang--Mills equation and balanced equation. In \sref{s:nonspur}, we construct an extension $\cQ$ and $\cDb$--operator whose cohomology describe the F-terms and the harmonic representatives the D-terms.

\section{Review of results}
\subsection{$\ap$--corrected Supergravity}
The heterotic action is fixed by supersymmetry up to and including $\ap^2$ corrections. There is nice basis of fields in which the action, equations of motion and supersymmetry variations are particularly compact. This was constructed  in \cite{Bergshoeff:1988nn} to order $\ap$  by supersymmetrising the Lorentz--Chern--Simons terms in $H$ and extended to $\ap^2$ in \cite{Bergshoeff:1989de}. The action and Bianchi identity for $H$ is unique up to field redefinitions; there is no ambiguity in the theory.  The action, in the field basis of \cite{Anguelova:2010ed}, is given by 
\begin{equation}
S = \frac{1}{2\kappa_{10}^2} \int\! \dd^{10\,}\! X \sqrt{g_{10}}\, e^{-2\Phi} \Big\{ \cR -
\half |H|^2  + 4(\del \Phi)^2 - \frac{\alpha'}{4}\big( \tr |F|^2 {-} \tr |R(\Theta^+)|^2 \big) \Big\} + \cO(\alpha'^3)~.
\label{eq:10daction}
\end{equation}
Our notation is such that  $\m,\n,\ldots$ are holomorphic indices along $\ccX$ with coordinates $x$; $m,n,\ldots$ are real indices along $\ccX$.   The 10D Newton constant is denoted by $\kappa_{10}$,
\hbox{$g_{10}=-\det(g_{MN})$}, $\Phi$ is the 10D dilaton, $\cR$ is the Ricci scalar evaluted using the Levi-Civita connection and $F$ is the Yang--Mills field strength with the trace taken in the adjoint of the gauge group. 

We take the $p$-form norm as $|T|^2 = \frac{1}{p!} T_{M_1 \cdots M_p} T^{M_1 \cdots M_p}$
and  the curvature squared terms correspond to
$$
\tr |F|^2 = \half \tr F_{MN} F^{MN}~~~\text{and}~~~ \tr |R(\Th^+)|^2 ~=~
\half  R_{MN}{}^A{}_B(\Theta^+) R^{MN}{}^B{}_A(\Theta^+)~,
$$
where the Riemann curvature is evaluated using a twisted connection
$$
\Theta^\pm_M{}^A{}_B = \Theta_M{}^A{}_B \pm \half H_M{}^A{}_B~,
$$
with $\Theta_M$ is the Levi-Civita connection and $A,B$ are the tangent space indices. The three--form $H$ satisfies a Bianchi--identity
\beq
\dd H \=\!- \frac{\ap}{4} \left( \tr F^2 - \tr R^2\, \right)~,
\label{eq:Anomaly0}\eeq
while at the same time it is related to the hermitian form as $H = \dd^c \o$. With respect to a fixed complex structure this corresponds to
$$
H = \ii (\del - \delb) \o~,
$$ 
and so the Bianchi identity in this complex structure is
\beq\label{eq:Bianchi2}
2\ii \del \delb \o =  \frac{\ap}{4} \left( \tr F^2 - \tr R^2\, \right)~.
\eeq 
 
 The decomposition of the connections into type is
$$
A \= \A - \A^\dag~, \qquad \Th \= \th - \th^\dag~.
$$
where $\A = A^{(0,1)}$ and $\th = \Th^{(0,1)}$ and we are using conventions in which $A$ and $\Th$ are antihermitian.

All we care about is that the  Bianchi identity \eqref{eq:Anomaly0}, equations of motion  and  supersymmetry variations  match calculations from string scattering amplitudes \cite{Metsaev:1987zx} which has been checked \cite{Chemissany:2007he}. This data is fixed, up to the usual caveat of field redefinitions. Indeed, if one were to perform a field redefinition, so that for example the curvature tensor $R$ in the Bianchi identity is evaluated with a different connection, then the field redefinitions will propagate through the supersymmetry variations and equations of motion, likely losing their simple closed form. For example, if one evaluated $R$ using the Chern connection  then this requires a field redefinition in which resulting in a metric $\wt g$ that is no longer a tensor -- it is charged under gauge transformations like the B-field -- and this field redefinition is likely modify the supersymmetry variations already at order $\ap$.\footnote{This is nicely described in \cite{Melnikov:2014ywa}. If one wishes to preserve manifest $(0,2)$--supersymmetry on the worldsheet, then $R$ in the Bianchi identity should be evaluated with the Chern connection. If one wishes to preserve manifest $(0,1)$--supersymmetry  with $g$ a conventional metric tensor then one evaluates the Bianchi identity with $\Th^+$. } We choose to use the usual conventions in which $g$ is a metric tensor and $R$ is evaluated with $\Th^+$.

From now on we work to first order in $\ap$. For emphasis, we will sometimes include $+\;\cO(\alpha'^2)$ in our equations, but this will mostly be suppressed. 

\subsection{Holomorphic gauge and F-term equations}
 In the notation of \cite{Candelas:2016usb,Candelas:2018lib} the  field deformations, parameterised by the real coordinate $y^a$,  obey the following equations to first order in $\ap$:
\beq\label{eq:moduliEqnHol}
\begin{split}
\delb \D_a{}^\m &\= 0~, \\[0.15cm]
\delb_\A (\fD_a \A) &\= \D_a{}^\m F_\m~, \\
\del \ccZ_a^{(0,2)} + \delb \ccZ_a^{(1,1)} &\= 2\ii \, \D_a{}^\m (\del\o)_\m +\frac{\ap}{2} \tr \big( \fD_a \A \, F\big) - \frac{\ap}{2} \tr \big( \fD_a \Th \, R\big) ~,\\
 \delb \ccZ_a^{(0,2)} &\= 0~.
\end{split}
\eeq
We assume  $h^{(0,2)}=0$ and so the last line is $\ccZ_a^{(0,2)} = \delb \b_a^{(0,1)}$ and this means left hand side of the third line is $\delb$--exact. 

There are equations coming from the HYM equation $\o^2 F = 0$ and the balanced equation  $\dd (\o^2) = 0$ which are analysed below in \sref{s:dterm}. Both these equations and \eqref{eq:moduliEqnHol} are invariant under small gauge transformations \citeSG:
\beq\label{eq:smallTransf}
\begin{split}
 \D_a{}^\m &\sim \D_a{}^\m + \delb \ve_a{}^\m\,,   \quad \fD_a \A \sim \fD_a \A + \ve_a{}^\m F_\m + \delb_\A \phi_a\,, \\[7pt]
 \ccZ_a &\sim \ccZ_a  + \ve_a{}^m (H+\ii \dd \o)_m + \frac{\ap}{2} \tr\! \left( F \phi_a \right)  + \dd\!\left (\mfb_a + \ii \ve_a{}^m \o_m\right)~,\\[5pt]
 \ccZb_a &\sim \ccZb_a  + \ve_a{}^m (H-\ii \dd \o)_m + \frac{\ap}{2} \tr\! \left( F \phi_a \right)+ \dd\! \left(\mfb_a - \ii \ve_a{}^m \o_m\right)~,\\
\end{split}\raisetag{1.4cm}
\eeq
where $\ccZ_a = \ccB_a + \ii \del_a \o$ and $\ccZb_a = \ccB_a - \ii \del_a \o$. 
A convenient choice of gauge fixing is holomorphic gauge:
\beq\label{eq:finalGauge}
\begin{split}
   \D_\ab{}^\m &\= 0~, \quad \d \O^{(3,0)} \= \d y^\a k_\a \O~, \quad     \fD_\ab \A \= 0~,\\
    \quad  \ccZ_\ab^{(1,1)} &\= 0~, \quad \ccZ_\ab^{(0,2)} \= \ccZb_\ab^{(0,2)} \= \ccZb_\ab^{(2,0)} \= 0~.\\
\end{split}
\eeq
In this gauge, we have, for example,
\beq\label{eq:modulieqnHolGauge}
\begin{split}
\delb \D_{\a}{}^\m &\= 0~, \qquad
\delb_\A (\fD_\a \A) + F_\m  \D_\a{}^\m \= 0~, \\
\end{split}
\eeq
where $\D_\a{}^\m$ is a $(0,1)$--form valued in $\ccT_\ccX^{(1,0)}$ and $\fD_\ab \A = 0$. The gauge simplifies other equations such as
\beq\label{eq:ccZeomgauge}
 \delb \ccZ_\a^{(1,1)}  - 2\ii \, \D_\a{}^\m (\del\o)_\m + \frac{\ap}{2} \tr \big( R \,  \fD_\a \Th\big) -\frac{\ap}{2} \tr \big( F\, \fD_\a \A \big)  \= 0~,
\eeq
and determines 
 $$
 \ccZ_\a^{(1,1)} \= 2\ii \fD_\a \o^{(1,1)} = 2 \ccB_\a^{(1,1)}~, \qquad \ccZb_\a^{(0,2)} {\=}2 \ccB_\a^{(0,2)} {\=}\! -2\ii \ccD_\a \o^{(0,2)}~.
 $$
While the field $\ccZ_\a^{(0,2)}$ vanishes in this gauge fixing, the field $\ccZb_\a^{(0,2)}$ does not. Hence, it is a physical degree of freedom whose role in a heterotic vacuum is not yet clear.  The role of $\ccZ_\a^{(1,1)}$ is analogous to the complexified \K modulus for a CY manifold. 

We refer to \eqref{eq:modulieqnHolGauge} and \eqref{eq:ccZeomgauge} as F-term equations. 
\subsection{D-term equations}\label{s:dterm}
An analysis of deformations in holomorphic gauge of the balanced equation and hermitian Yang--Mills equation was presented in \citeSG. This gives a relation between deformations in terms of adjoint operators. In this section we revist this calculation massaging the equations with a prescenice of results to come. 

The top--form $\O$ on $X$ has a norm
$$
\norm{\O} \= \frac{1}{3!} \O_{mnp} \Ob^{mnp}~.
$$
It is related to the dilaton $\dd \log \norm{\O} = \! -2 \dd \phi$.
We gauge fix as in \cite{Anguelova:2010ed,McOrist:2019mxh} so that the dilaton is a constant on $X$ and so in this gauge  $\norm{\O}$ is also a constant. Furthermore, in holomorphic gauge \eqref{eq:finalGauge},  $(\del_\a \O)^{(3,0)} = k_\a \O $, where $k_\a$ is a constant on $X$. Using this,  a first order variation of the norm is
$$
\del_\a \norm{\O}^2 \=\norm{\O}^2  (k_\a - \del_\a \log \sqrt{g})~.
$$
As $\norm{\O}$ and $k_\a$ are constants over $X$ and it follows 
$ \o^{\l\sb} \del_\a \o_{\l\sb} =\frac{1}{2\ii} \o^{\l\sb} \ccZ_{\a\,\l\sb} $ is a constant on $X$. 

A first order holomorphic variation of the balanced equation is
\begin{align}
\partial(\omega\,\ccZ_\a^{(1,1)})&\=0~,\\
-i\delb(\omega\,\ccZ_\a^{(1,1)})+2\,\del(\omega\,{\Delta_\a}^\mu\omega_\mu)&\=0~.
\label{eq:defbal2}
\end{align}
Using the Hodge dual relation for a $(1,1)$--form  \eqref{eq:Hodge2form}, together with  $\o^{\l\sb} \ccZ_{\a\,\l\sb}$ being a constant and the balanced equation, the first equation amounts to 
\begin{equation}
\label{eq:coclosedZ}
\delb^\dagger\ccZ_\a^{(1,1)}=0\:.
\end{equation}
The second equation is
\beq
\del (\del_\a \o^{(0,2)} ) \o + \delb (\del_\a \o^{(1,1)})  \o + (\del_\a \o^{(0,2)} ) \del \o + (\del_\a \o^{(1,1)}) \delb \o \= 0~.
\eeq
The Hodge dual of a $(2,3)$--form in \eqref{eq:Hodge23} can be used together with the gauge fixed F-term equation \eqref{eq:ccZeomgauge} and $\nabla^{{\rm Ch/H}}_\m \Delta_{\a}{}^\m = 0$ as well as the background equations of motion $H_{\m\n}{}^\n = 0$, $\o^2 F= 0$, $\del_\a \o_{\mb\nb} = 2\ii  \D_{\a[\mb\nb]}$ to rewrite this equation in terms of contractions. This result is
\begin{equation}\label{eq:adjoint1}
\frac{i}{2}\,(\delb\omega)^{\n\r\mb}\ccZ_{\a\,\r\mb} -\frac{\ap}{4}\Big[\tr\big(F^{\nu\mb} \,\fD_\a \A_{\mb}\big) - \tr\big(R^{\n\mb} \fD_\a \Th_{\mb}\big) \Big] + \,\nabla^{{\rm Ch/H}\, \mb}\Delta_{\a\mb}{}^\n + \ii \D_{\a\,\rb\lb} (\del \o)^{\n\rb\lb}\=0~,
\end{equation}
where $\nabla^{{\rm Ch/H}\, \mb}$ is the Chern or Hull connection (see \citeSG for discussion on this ambiguity) and because $H = \cO(\ap)$ and $\D_{\a[\rb\lb]} = \cO(\ap)$ the last term is $\cO(\ap^2)$ and so is dropped from hereon. A good consistency check is to derive this equation by differentiating $g^{\m\nb} H_{\m\nb\rb} = 0$. 

A first order variation of the Hermitian Yang-Mills equation can be written as 
\beq
\label{eq:adjoint2}
\delb_\A^\dag (\fD_\a \A )+\half F^{\m\nb} \ccZ_{\a\,\m\nb}     \= 0~,
\eeq
A similar equation is satisfied for the connection $\Th$ on the tangent bundle  to this order in $\ap$. That is
\beq
\label{eq:adjoint3}
\delb_\Th^\dag (\fD_\a \Th )+\half R^{\m\nb} \ccZ_{\a\,\m\nb}     \= 0~.
\eeq

In holomorphic gauge, we declare `F-term' type equations to be \eqref{eq:modulieqnHolGauge}--\eqref{eq:ccZeomgauge}. We do not have a good definition of string field theory, so we use this terminology with care. The motivation for the nomenclature is that these are the equations, together with an appropriate definition of holomorphy, that derive from a superpotential type construction. The equations \eqref{eq:adjoint1}--\eqref{eq:adjoint3} deriving from the balanced and HYM equation we declare to be D-term equations. A main outcome of this paper is to show that there is a $\cDb$--operator acting on first order fluctuations whose kernel corresponds to F-term equations, and co--kernel to the D-term equations. This is helps justifies this terminology. 

\subsection{The Hodge decomposition}
The balanced and HYM equations play a role in determining the exact and co--exact terms in the Hodge decomopsition of fields. We present this to emphasise the fact that deformations of a heterotic theory are not simply the harmonic representatives. Taking into account the F-term equations, the Hodge decomposition is
\beq\label{eq:HodgeHet1}
\begin{split}
 \ccZ_\a^{(1,1)} &\= \ccZ_\a^{(1,1)\,harm}   +  \delb^\dag\xi_\a^{(1,2)}~,\\
\D_\a &\= \D_\a{}^{harm} + \delb \k_\a  ~,\\
 \fD_\a \A &\= \fD_\a \A^{harm} + \delb_\A  \Phi_\a + \delb_\A^\dag \Psi_\a^{(0,2)}~.\\
\end{split}
\eeq
The role of the cohomology we aim to compute is to tell us what linear combinations of harmonic terms in \eqref{eq:HodgeHet1} correspond to unobstructed deformations. The exact and co--exact terms in the Hodge decomposition are physical and so are important and are determined by substituting into the HYM and balanced equations and solving the corresponding Poission equations:
\begin{subequations}
\begin{align}
 \xi_\a^{(1,2)} &\= \Box_\delb^{-1} \Big(  \D_\a{}^\m (\del\o)_\m - \frac{\ii\ap}{4} \tr ( \fD_\a \A F )\notag\\
 &\hspace{2.5cm} + \frac{\ii\ap}{2} \, \big( \nabla_\n \D_\a{}^\m + \half \, \nabla^\m \, \ccZ_{\a\,\n}^{(0,1)} \big) R^\n{}_\m \Big)  + \delb^\dag\text{--\,closed} ~,\label{eq:HodgeHeterotic2a}\\[0.1cm]
 \k_\a{}^\m  &\= \Box_{\delb}^{-1}\Big( - \half \, \big( \del^\dag \ccZ_\a \big)^\m + \frac{1}{4\ii}( \ccZ_{\a\,\nb\rb})(\del\o)^{\nb\rb\m}\Big)~,\label{eq:HodgeHeterotic2b}\\[0.2cm]
 \Psi_\a^{(0,2)} &\=  \Box^{-1}_{\delb_\A} ( \D_\a{}^\m F_\m) + \delb_\A^\dag\text{--\,closed} ~, \qquad \Phi_\a \= - \half \Box_{\delb_\A}^{-1} \Big( (\ccZ_\a^{\m\nb}) F_{\m\nb}\Big)~.\label{eq:HodgeHeterotic2c}
 \end{align}
\end{subequations}
The point here is that  deformations of fields are not only highly coupled but they are not harmonic representatives. A natural question to ask is: can I change the gauge to find a harmonic decomposition? The answer is not likely without significant sacrifice. For example, one easily loses  the holomorphic dependence of deformations on parameters. This is unlike the study of CY manifolds in which one studies  holomorphic deformations that are also harmonic, viz. $\delb$--harmonic $(1,1)$ forms and $(2,1)$--forms. This derives from $H=0$, an additional constraint that we do not have in the more general situation. 

One point of this paper is to point out that without sacrificing holomorphy, the complicated Hodge decomposition beautifully reorganises itself into harmonic representatives of a $\cDb$--operator on a certain extension bundle.

\section{Warm-up: Extension bundles for complex manifolds}
\label{s:atiyah}
One approach to understanding \eqref{eq:moduliEqnHol}--\eqref{eq:finalGauge} is inspired from the work of Atiyah \cite{Atiyah:1955} who described the deformations of a holomorphic bundle on a complex manifold. Its instructive to review this with an eye towards the full heterotic story.

Consider a holomorphic bundle on a complex manifold. Atiyah, in 1955, pointed out that not all complex structure deformations are allowed as some may introduce a non--trivial $F^{(0,2)}$ component destroying holomorphy of the bundle. Those that are allowed satisfy $$\delb_\A (\fD_\a \A) +   F_\m \D_\a{}^\m \= 0~, \quad \text{and} \quad \delb \D_\a \= 0.$$ Intuitively such complex structure deformations may introduce $F^{(0,2)}$ but can compensated by a simultaneous deformation of the bundle. A way to realise these constraints is to define a vector bundle $Q_1$ as a short exact sequence
\beq\label{eq:Q1def}
\xymatrix{
0 \ar@{>}[r] &\EndE  \ar@{>}[r]^{i_1}  & Q_1 \ar@{>}[r]^{\pi_1}& \ccT^{(1,0)}_\ccX   \ar@{>}[r] & 0}~,
\eeq
where $\pi_1: Q_1 \to \ccT_\ccX$ is the canonical projection and $i_1$ the inclusion map.\footnote{The sequence is exact so $\ker{\pi_1} = \im i_1 = \EndE$ and by rank--nullity the dimension of a typical fiber at $Q_1$  is $\dim (\EndE) + \dim (\ccT_\ccX)$. } One can then study $(0,p)$--forms valued in $Q$. These are vectors, schematically
$$
\begin{pmatrix}
 \d_\a \A^{(p)} \\
 \D_\a^{(p)} 
\end{pmatrix}~,
$$
where $\d_\a \A^{(p)}$ is a $(0,p)$--form valued in ${\rm End}\,E$ and $\D_\a^{(p)}$ is a $(0,p)$--form valued in $\ccT_\ccX^{(1,0)}$. 

There is a holomorphic structure defined by an operator $\Db_1$ given by 
$$
\Db_1 \= 
\begin{pmatrix}
 \delb_\A & \cF \\
 0 & \delb
\end{pmatrix}~, \qquad \cF: \O^{(0,p)}(\ccX,\ccT_\ccX^{(1,0)}) \to \O^{(0,p+1)}(\ccX, \EndE) ~, \quad \cF(\D^{(p)}_\a) \= F_\m \D^{(p)}_\a{}^\m ~,
$$ 
where, for example, $\O^{(p,q)}(\ccX, \EndE)$ denotes $(p,q)$--forms on $\ccX$ valued in $\EndE$.
The action of $\Db_1$ on a typical fibre of $Q_1$ vanishes if the field deformations obey the requisite equations of motion:
$$
\Db_1 \begin{pmatrix}
 \d_\a \A^{(p)} \\
 \D^{(p)}_\a
\end{pmatrix} \=  \begin{pmatrix}
\delb_\A (\d_\a \A^{(p)})  + \cF(\D^{(p)}_\a) \\
\delb \D^{(p)}_\a
\end{pmatrix} \= 0
~.
$$ 
Furthermore, $\Db_1^2 = 0$ if and only if the $F^{(0,2)} = 0$ and its Bianchi identity holds $\dd_A F = 0$. 

 Under a  small gauge transformation, which includes a small diffeomorphisms for generality,
\beq\label{eq:smallgaugebundle}
\begin{split}
 \d_\a \A^{(p)} &\longrightarrow \d_\a \A^{(p)} + \ve_\a{}^\m F_\m + \delb_\A \phi_\a~, \quad \d_\a \A^{(p)\,\dag} \longrightarrow \d_\a \A^{(p)\,\dag} + \ve_\a{}^\mb F_\mb - \del_{\A^\dag} \phi_\a~,\\[4pt]
   \D^{(p)}_\a{}^\m &\longrightarrow \D^{(p)}_\a{}^\m + \delb \ve_\a{}^\m~,\qquad\qquad\qquad  \D^{(p)}_\a{}^\mb \longrightarrow \D^{(p)}_\a{}^\mb + \delb \ve_\a{}^\mb~,
\end{split}
\eeq
where $\ve_\a{}^m$ is a vector valued $(0,p-1)$--form parameterising the small diffeomorphism, and $\phi_\a$ is a $(0,p-1)$--form valued in $\EndE$ parameterising the small gauge transformation.   

The bundle $Q$ and operator $\Db_1$ are defined with respect to holomorphic deformations only. The operator $\Db_1$ is equivariant with respect to gauge transformations:
$$
\Db_1 \begin{pmatrix}
 \d_\a \A^{(p)} + \ve_\a{}^\m F_\m + \delb_\A \phi_\a  \\
 \D^{(p)}_\a + \delb \ve_\a{}^\m
\end{pmatrix} \= \Db_1 \begin{pmatrix}
 \d_\a \A^{(p)}   \\
 \D_\a^{(p)}  
 \end{pmatrix}
~.
$$ 
For $p=1$ we enter the case of interest, which are deformations of a complex manifold with bundle. The  notation for deformations $\d_\a \A^{(1)} \cong \fD_\a \A$ and $\D_\a^{(1)} \cong \D_\a$.  The space of deformations is the cohomology $H^{(0,1)}(\ccX, Q_1)$.  However, as can be seen from \eqref{eq:smallgaugebundle} together with the bundle being stable, so $\delb_\A \phi_\a = 0$ has only trivial solutions, any gauge transformation will introduce a non--holomorphic dependence on parameters. i.e. $\fD_\a \A^\dag \ne 0$.  This gauge choice is already made in many physical calculations such as  the moduli space metric \citeM or results deriving from the superpotential, e.g. \citeS. So the cohomology is to be treated with care: we are always gauge fixed, and so always fixed a representative. This becomes more important in the situation to come. \footnote{The Kaluza--Klein reduction of heterotic supergravity at the standard embedding is almost always performed in harmonic gauge. Deformations of the complexified \K form and deformations of complex structure are then zero modes of Laplacians. Quantities such as the moduli space metric \cite{Candelas:1989bb} are calculated with this particular choice of representative. Interestingly, this moduli space metric is not manifestly gauge invariant. Also note that the standard embedding holomorphic gauge and harmonic gauge are the same. Outside the standard embedding, holomorphic gauge is preferred as it naturally connects the complex structure of $X$ with the moduli space $\cM$. }

  Setting this issue aside one can use a long exact sequence in cohomologies inherited from the short exact sequence to show 
 \beq\label{eq:AtiyahCohomology1}
  H^{(0,1)}(\ccX, Q_1) \= H^{(0,1)}_{\delb_\A} (\ccX, \EndE) \oplus \ker \cF~.
 \eeq
 The Hodge decomposition of an  element of $H^1_{\Db_1}(\ccX,Q_1)$ is
$$
\D_\a \= \D_\a{}^{harm} + \delb \k_\a  ~,\qquad
 \fD_\a \A \= \fD_\a \A^{harm} + \delb_\A  \Phi_\a + \delb_\A^\dag \Psi_\a^{(0,2)}~.\\
$$
Equation \eqref{eq:AtiyahCohomology1} implies the tangent space to the moduli space is a direct sum. So there is a choice of coordinates for the moduli space in which parameters decompose into two components  $y^\a = (y^{\a_1}, y^{\a_2})$.  The first component is one-to-one with $H^1(\ccX,\EndE)$: 
$$
\D_{\a_1} \= \delb \k_{\a_1}  ~,\qquad
 \fD_{\a_1} \A \= \fD_{\a_1} \A^{harm} + \delb_\A  \Phi_{\a_1}~.\\
$$
where we use $\delb_\A (\fD_{\a_1} \A )= 0$ and so $\D_{\a_1}{}^\m F_\m= 0$. This implies $\k_{\a_1}{}^\m F_\m = \delb_\A(\cdots)$, which has a solution provided we can invert $F_\m$ and solve for $\k_{\a_1}{}^\m$. The terms $\k_{\a_1}$ and $\Phi_{\a_1}$ are completely undetermined at this point and so there is an infinite dimensional space of solutions. In the full heterotic theory, there are additional equations of motion, and these fix $\k_{\a_1}$ and $\Phi_{\a_1}$. 

The second component of solutions is one-to-one with $\ker \cF \subset H^1(\ccX,\ccT_\ccX^{(1,0)})$:
$$
\d y^{\a_2} \D_{\a_2} \= \d y^{\a_2}  \left( \D_{\a_2}^{harm} + \delb \k_{\a_2}\right)  ~,\qquad
 \d y^{\a_2} \fD_{\a_2} \A \= \d y^{\a_2} \left( \delb_\A  \Phi_{\a_2} + \delb_\A^\dag \Psi^{(0,2)}_{\a_2}\right)~,\\
$$
where $\d y^{\a_2}$ is a vector in $\ker \cF$. That is, the cohomology tells us what harmonic representative of $H^1(\ccX,\ccT_\ccX^{(1,0)})$ appear in this equation. The Atiyah equation implies $\Psi^{(0,2)}_{\a_2} = \Box^{-1}_{\delb_\A} (\D_{\a_2}{}^\m F_\m)$ while  $\k_{\a_2}$ and $\Phi_{\a_2}$ are undetermined.  

The lesson here is that in writing $  H^{(0,1)}(\ccX, Q_1) \= H^{(0,1)}_{\delb_\A} (\ccX, \EndE) \oplus \ker \cF$ we do not mean that the field deformations decompose into a direct product. Instead, it is telling us about what combinations of harmonic forms can appear and that they have a Hodge decomposition. This example, we have some undetermined exact terms; in the full heterotic theory these are all fixed by the equations of motion and gauge fixing.

\section{Heterotic theories with spurious degrees of freedom}
\label{s:spur}
\subsection{F-terms}
\label{s:spurfterm}
The situation in the Hull--Strominger system is more complicated as we have hermitian deformations, the B-field and a set of equations in \eqref{eq:moduliEqnHol}--\eqref{eq:finalGauge} together with a Bianchi identity. The approach of \citeOS is to include deformations of the tangent bundle $\d \Th^{(0,1)} = \d y^\a \fD_\a \th$ as independent degrees of freedom. The tangent bundle $\ccT_\ccX$ is holomorphic  with $R^{(0,2)}=0$, and the curvature two--form obeys the HYM equation $\o^2 R = 0$.\footnote{If you assume the Bianchi identity and supersymmetry variations hold then equations of motion, to first order in $\ap$, hold if and only if $\o^2 R = 0$. This result is valid to first order in $\ap$ as originally derived in \cite{Hull:1986kz,Ivanov:2009rh} with an elegant summary of the result in the appendix of \cite{Martelli:2010jx}.} Hence, $\fD_\a \th$ obeys an Atiyah equation, of the same form as the second equation of \eqref{eq:moduliEqnHol}. These we refer to as spurious degrees of freedom because  in the physical theory the deformations $\fD_\a \th$ are determined in terms of the other deformations $\D_\a$, $\fD_\a \A$ and $\fD_\a \o$. Nonetheless, they lead to a nice mathematical result which we review and refine here. Define an extension $Q_2$ of $Q_1$:
\beq
\xymatrix{
0 \ar@{>}[r] &{\rm End}(\ccT_\ccX)  \ar@{>}[r]^{i_2}  & Q_2 \ar@{>}[r]^{\pi_2}& Q_1   \ar@{>}[r] & 0}~,
\notag\eeq
with a differential operator
$$
\Db_2 \= 
\begin{pmatrix}
 \delb_\th & 0 & \cR\\
 0 & \delb_\A & \cF\\
 0 & 0 & \delb 
\end{pmatrix}~,
$$
which acts on forms valued in $Q_2$. The operator $\cR$ is analogous to $\cF$: $\cR(\D_\a) =  R_\m \D_\a{}^\m$ where $R_\m$ is the curvature two--form for $\th$. The operator $\Db_2^2 = 0$ provided $R^{(0,2)} = 0$ and its Bianchi identity holds.

To incorporate hermitian deformations define a bundle $Q$ by the short exact sequence
\beq
\xymatrix{
0 \ar@{>}[r] &\ccT^{(0,1)}_\ccX  \ar@{>}[r]^{i}  & Q \ar@{>}[r]^{\pi}& Q_2   \ar@{>}[r] & 0}~.
\notag\eeq
We consider $(0,p)$--forms valued in $Q$ denoted as
 $$
 \ccY^{(p)}_{\a}\= 
\begin{pmatrix}
 \ccZ^{(p)}_{\a\,\n} \\
\d_\a \th^{(p)} \\
\d_\a \A^{(p)} \\
 \D^{(p)}_\a{}
\end{pmatrix}~.
 $$
The first row is a $(0,p)$--form valued in $\ccT_\ccX^{*\,(1,0)}$, the middle two rows are $(0,p)$--forms valued in ${\rm End} \ccT_\ccX$ and ${\rm End} \ccE$ respectively, while the last line is a $(0,p)$--form valued in $\ccT_\ccX^{(1,0)}$. The special case of $\ccY_\a^{(1)}$ corresponds to the field deformations of interest, where $\ccZ_{\a\,\n}^{(1)} \cong \ccZ_{\a\,\n\mb}\dd x^\mb$, $\d_\a \A^{(1)} \cong \fD_\a \A$ and $\d_\a \th^{(1)} \cong \fD_\a \th$. We will need this more general construction in later sections and when we construct the adjoint operator.

In the appendix \sref{s:Dbops} we find a family of operators that satisfy $\Db^2 = 0$ off-shell and reproduce the F-term equations. The operator in \citeOS is an example of this family. In this paper we use a $\Db$--operator in this family whose presentation is simpler:
$$
\Db \= 
\begin{pmatrix}
-\delb & \cH \\
0 & \Db_2
\end{pmatrix}~,
$$
where $\cH:\O^{(0,p)}(\ccX,Q_2) \to \O^{(0,p+1)}(\ccX, \ccT_\ccX^{*\,(1,0)})$ is a linear operator
\beq
\cH_\n (\d_\a \th^{(p)},  \d_\a \A^{(p)}, \D_\a^{(p)} ) \=  2\ii  \D_\a^{(p)\,\m}(\del\o)_{\m\n} + \frac{\ap}{2}  \left( \tr (\d_\a \A^{(p)} F_\n) - \tr (\d_\a \th^{(p)} R_\n) \right)~,\notag\eeq
and we show in \sref{s:DbsqSpur} that $\Db^2 = 0$ off-shell. 
We see that 
$$
\Db \ccY_\a^{(1)} \= 0~,
$$
 corresponds to the F-term equations \eqref{eq:modulieqnHolGauge}--\eqref{eq:ccZeomgauge}.

We note that 
$$ \cH_\n\Big( \Db_2 (\d_\a \th^{(p)} ,  \d_\a \A^{(p)}, \D_\a^{(p)} )\Big) = \delb \cH_\n(\d_\a \th^{(p)} , \d_\a \A^{(p)}, \D^{(p)}_\a)~, 
$$ 
while under a small gauge transformation \eqref{eq:smallTransf}
\beq\label{eq:Hgauge}
\begin{split}
\cH_\n\Big(\d_\a \th^{(p)}{ + } \delb_\th \psi_\a {+} \ve_\a{}^\m R_\m,& \d_\a \A^{(p)} {+}  \delb_\A \phi_\a {+} \ve_\a{}^\m F_\m,  \D^{(p)}_\a {+} \delb \ve_\a \Big)  \\
&\=\cH_\n(\d_\a \th^{(p)}, \d_\a \A^{(p)}, \D_\a) +  \delb \cH_\n(\psi_\a,\phi_\a, \ve_\a{}^\m) ~.
\end{split}
\eeq
It is important that the gauge symmetry has an action on the complexified hermitian term
$$
\ccZ_\a^{(1,1)} \sim \ccZ_\a^{(1,1)} + 2\ii \ve_\a{}^\m (\del \o)_\m   + \frac{\ap}{2} \tr (\phi_\a F) + \del (\mathfrak{b}_\a^{(0,1)} + \ii \ve_\a{}^\m \o_\m) + \delb \mathfrak{b}_\a^{(1,0)} ~.
$$
Here $\mathfrak{b}_\a$ is a one--form associated to gauge transformations of the B-field. At first sight, we seem to have a problem: $\Db \ccY_\a^{(1)} = 0$  is no longer satisfied as  the term  ${\del (\mathfrak{b}_\a^{(0,1)} + \ii \ve_\a{}^\m \o_\m)}$ spoils the equation. The resolution is that \eqref{eq:ccZeomgauge} is derived from \eqref{eq:moduliEqnHol}  in holomorphic gauge. There are no residual gauge transformations that preserve holomorphic gauge which means any small gauge transformation is going to violate the gauge fixing condition. Instead of \eqref{eq:ccZeomgauge} we need to use is the third line of \eqref{eq:moduliEqnHol} and this involves $\ccZ^{(0,2)}_\a$, which is no longer zero. Indeed, we need to pair the transformation of $\ccZ_\a^{(1,1)}$ with
$$
\ccZ_\a^{(0,2)} \sim \ccZ_\a^{(0,2)} + \delb (\mathfrak{b}_\a^{(0,1)} + \ii \ve_\a{}^\m \o_\m)~.
$$

Then we find that under a small gauge transformation
$$
\delb \ccZ_\a^{(1,1)} + \del \ccZ_\a^{(0,2)} \sim  \delb \ccZ_\a^{(1,1)} + \del \ccZ_\a^{(0,2)} + \delb \ccH(\psi_\a,\phi_\a, \ve_\a{}^\m, )~.
$$

As $h^{(0,2)} = 0$, the equations of motion imply $\ccZ_\a^{(0,2)} = \delb \b_\a^{(0,1)}$:
$$
\delb \ccZ_\a^{(1,1)} + \del \ccZ_\a^{(0,2)} = \delb (\ccZ_\a^{(1,1)} - \del \b_\a^{(0,1)}) = \cH (\fD_\a \th, \fD_\a \A, \D_\a)~,
$$
where $\b_\a$ is a gauge dependent quantity. In holomorphic gauge it vanishes. A gauge invariant formulation likely involves this combination. It would be interesting to relate these observations to the complexified gauge transformations studied in \cite{Ashmore:2019rkx}, particularly in light of the D-term calculations we perform below. 

The lesson is that the cohomological property
$$
\cH: H^{(0,p)}_{\Db_1} (X, Q_1) \longrightarrow H_\delb^{(1,p+1)} (\ccX)~,
$$
 is subtle. A better understanding of complexified gauge symmetries and GIT would probably help.

Setting this issue aside,  it is shown in \citeOS that there is a long exact sequence in cohomology to describe first order deformations of $\Db$: 
\beq
H^1_{\raisebox{-2pt}{$\scriptstyle\overline D$}}(\ccX,Q) = H^1(\ccX,\ccT^{\raisebox{1pt}{$*$}}_\ccX) \oplus \ker\!\cH~,\qquad \ker\!\cH\subseteq H^1(\ccX, Q_2)~,
\label{AllModuli}\eeq
where
\beq
H^1(\ccX, Q_2) = H^1(\ccX, {\rm End}\,\ccT_\ccX)\oplus \Hend\oplus (\ker\!\cF\cap\ker\!\cR)~.
\label{HolModuli}\eeq
There are the hermitian Yang--Mills and balanced equations to take into account. We show their role appears in the adjoint operator $\Db^\dag$.

\subsection{The D-terms and the adjoint operator $\overline{D}^\dag$}
\label{s:spuradjoint}
 A metric on $Q$ describes how to pair field deformations to produce a real number. In string theory, a natural metric to use is the moduli space metric. As we preserve $N=1$ supersymmetry in spacetime, the metric is \K.  With the spurious degrees of freedom, this was first derived in  \citeM using a dimensional reduction of the $\ap$--corrected supergravity theory considered here. The result in our notation is 
\beq\label{eq:modulimetricspurious}
\begin{split}
 g\#_{\a\bb} &\= \frac{1}{V}\int_X \Big( \frac{1}{4} \ccZ_\a^{(1,1)} \star\ccZb_\bb^{(1,1)}   +\frac{\ap}{4} \tr{ ( \fD_\a \th \star \fD_{\bb}\th ) } -  \frac{\ap}{4}\tr{ ( \fD_\a \A \star \fD_{\bb}\A^\dag ) }+ \D_\a{}^\m\star\D_{\bb}{}^{\nb}\,g_{\m\nb}  \Big)~,
\end{split}\eeq
 We have written the antihermitian connections as  $A = \A-\A^\dag$, $\A= A^{(0,1)}$ and  $\fD_\bb A = -\fD_\bb \A^\dag$ with similar expressions for $\Th = \th - \th^\dag$.

We take the metric on a pair of sections $\ccY_\a^{(p)}$ and $\ccY_\b^{(p)}$ to be the natural generalisation of the moduli space metric:
\beq\label{eq:bigmodulimetricspurious}
\begin{split}
 \langle\ccY^{(p)}_\a , \ccY^{(p)}_\b \rangle &\=  \frac{1}{V}\int_X \left(  \frac{1}{4} \ccZ_{\a\,\n}^{(p)} \star\ccZb_{\bb\,\mb}^{(p)} \, g^{\n\mb}  + \frac{\ap}{4} \tr{ ( \d_\a \th^{(p)} \star \d_{\bb}\th^{(p)}{}^\dag ) } \right.\\[3pt]
&\qquad\left. - \frac{\ap}{4}\tr{ ( \d_\a \A^{(p)} \star \d_{\bb}\A^{(p)}{}^\dag} ) + \D^{(p)}_\a{}^\m\star\D^{(p)}_{\bb}{}^{\nb}\,g_{\m\nb} \right)~,
\end{split}
\eeq
where we hermitian conjugate in the appropriate way so the metric is real. 

The metric allows us to define an adjoint operator $\Db^\dag$. 
\beq
\langle \Db \ccY_\a^{(p)}, \ccY_\b^{(p+1)} \rangle \= \langle \ccY_\a^{(p)} , \Db^\dag \ccY_\b^{(p+1)} \rangle~.
\eeq
It is instructive to explicitly compute:
\beq	\notag
\begin{split}
 \langle \Db \ccY_\a^{(p)}, \ccY_\b^{(p+1)} \rangle &\= \frac{1}{4V} \int_\ccX \Big( -\delb \ccZ_{\a\,\n}^{(p)} + 2\ii \,  \D^{(p)}_\a{}^\m (\del\o)_{\m\n} - \frac{\ap}{2} \tr \big(     \d_\a \th^{(p)} R_\n \big) + \frac{\ap}{2} \tr \big(    \d_\a \A^{(p)} \, F_\n \big)    \Big) \star \ccZb_{\bb\,\lb}^{(p+1)} g^{\n\lb}\\
 &\qquad\qquad + \frac{\ap}{4V}\tr \int  (\delb_\th \,(\d_\a \th^{(p)}) + R_\m \D^{(p)}_\a{}^\m) \star \d_\bb \th^{\dag\,(p+1)} \\
&\qquad\qquad - \frac{\ap}{4V}\tr \int  (\delb_\A (\d_\a \A^{(p)}) +F_\m \D^{(p)}_\a{}^\m ) \star \d_\bb \A^{(p+1)\,\dag}+\frac{1}{V} \int \delb \D^{(p)}_\a{}^\m \star \D^{(p+1)}_\bb{}^\nb g_{\m\nb}\\
&\= \frac{1}{V} \int_\ccX \left\{ \frac{1}{4} \ccZ_{\a}^{(p)} \star (- \del^\dag \ccZb_\bb^{(p+1)}) 
+ \frac{\ap}{4} \tr \left( \d_\a \th^{(p)} \star \left((-1)^{p+1}\half R_{\m\nb} \ccZb_\bb{}^{\m\nb\,(p+1)} + \del^\dag_{\th^\dag} \d_\bb \th^{\dag\,(p+1)} \right) \right)\right.\\
&\qquad\qquad - \frac{\ap}{4} \tr \left( \d_\a \A^{(p)} \star \left((-1)^{p+1}\half F_{\n\sb} \ccZb_\bb{}^{\n\sb\,(p+1)} + \del^\dag_{\A^\dag} \d_\bb \A^{\dag\,(p+1)} \right) \right)\\
&\!\!\!\!\!\!\!\!+\left. \D^{(p)}_\a{}^\m \star \left( (-1)^p \frac{\ii}{2} (\del\o)_\m{}^{\r\nb} \ccZb_{\bb\,\r\nb}{}^{(p+1)} {+} \frac{\ap}{4}  \tr R_{\m\nb}\, \d_\bb \th^{\dag\,\nb}{}^{\,(p+1)} {-}  \frac{\ap}{4}  \tr F_{\m\nb}\, \d_\bb \A^{\dag\,\nb}{}^{(p+1)}  {+} \del^\dag \D^{(p+1)}_{\bb}{}_\m \right) \right\}\\
&\=  \langle  \ccY_\a^{(p)}, \Db^\dag\ccY_\b^{(p+1)} \rangle~,
\end{split}\label{eq:Ddagderivation}
\eeq
where our notation is, for example, $\ccZ_\a{}^{\m\nb\,(p+1)} = \frac{1}{p!}\ccZ_\a{}^{\m\nb}{}_{\lb_1\cdots\lb_p} \dd x^{\lb_1\cdots\lb_p}$.  
Comparing with \eqref{eq:bigmodulimetricspurious} we identify
\beq\label{eq:Dbardag}
\Db^\dag \= 
 \begin{pmatrix}
-\delb^\dag & 0 & 0 & 0\\
\wt R^\dag & \delb_\th^\dag & 0 & 0 \\
 \wt F^\dag & 0 & \delb_\A^\dag & 0\\
\cH^\dag &  \wh \cR^\dag & \wh \cF^\dag & \delb^\dag
\end{pmatrix}~,
\eeq
where
\beq
\begin{split}
 \wt \cR^\dag: \O^{(0,p+1)}(\ccT_\ccX^{*\,(1,0)}) \longrightarrow \O^{(0,p)}({\rm End}~\ccT_\ccX)~, \qquad & \wt R^\dag(\ccZ_{\a\,\n}^{(p+1)}) \= (-1)^{p}\half R^{\n\rb} \ccZ^{(p)}_{\a\,\n\rb}~,\\
 \wt \cF^\dag: \O^{(0,p+1)}(\ccT_\ccX^{(0,1)}) \longrightarrow \O^{(0,p)}({\rm End}~\ccT_\ccX)~, \qquad & \wt F^\dag(\ccZ^{(p+1)}_\a) = (-1)^{p}\half F^{\r\nb} \ccZ^{(p)}_{\a\,\r\nb}~,\\
 \wh \cR^\dag: \O^{(0,p+1)}({\rm End} \ccT_\ccX) \longrightarrow \O^{(0,p)}( \ccT_\ccX^{(0,1)} )~, \qquad & \wh \cR^\dag(\d_\a \th^{(p+1)}) \= -\frac{\ap}{4} \tr (R^{\n\rb} \,\d_\a \th^{(p)}_\rb) ~,\\
  \wh \cF^\dag: \O^{(0,p+1)}({\rm End} \ccE) \longrightarrow \O^{(0,p)}( \ccT_\ccX^{(0,1)} )~, \qquad &\wh \cF^\dag(\d_\a \A^{(p+1)}) \= \frac{\ap}{4} \tr (F^{\n\rb}\, \d_\a \A_\rb^{(p)}) ~,  \\
   \cH^\dag:  \O^{(0,p+1)}(\ccT_\ccX^{*\,(1,0)}) \longrightarrow \O^{(0,p)}(\ccT_\ccX^{(1,0)}) ~, \qquad & \cH^\dag(\ccZ_{\a\,\n}^{(p+1)}) \= (-1)^{p+1}  \frac{\ii}{2} (\delb\o)^{\n\r\lb} \ccZ_{\a\,\r\lb}^{(p+1)}~.
\end{split}
\eeq

The  action of $\Db^\dag$ on $\ccY_\a^{(1)}$ is
\beq\label{eq:Dbadjaction}
\Db^\dag 
\begin{pmatrix}
 \ccZ_{\a\,\n}^{(1)} \\
 \fD_\a \th \\
 \fD_\a A \\
 \D_\a{}^\m
\end{pmatrix}
\=
\begin{pmatrix}
\nabla^\mb \ccZ_{\a\,\n\mb}  \\
 \delb_\th^\dag \fD_\a \th +\half R^{\n\mb} \ccZ_{\a\,\n\mb}\\
  \delb_\A^\dag \fD_\a \A + \half F^{\n\mb} \ccZ_{\a\,\n\mb}\\
 - \frac{\ii}{2} (\delb\o)^{\m\r\lb} \ccZ_{\a\,\r\lb} - \frac{\ap}{4} \tr (R^{\m\rb} \,\fD_\a \th_\rb)+ \frac{\ap}{4} \tr (F^{\m\rb}\, \fD_\a \A_\rb) - \nabla^{{\rm Ch/H}\, \mb}\Delta_{\a\mb}{}^\n
\end{pmatrix}~,
\eeq
where we use that $R$ and $F$ are antihermitian and in the last line, we used the balanced equation to derive an identity
\beq\label{eq:codiffDelta}
 \delb^\dag \D^\m ~= - \star \frac{\ii \o^2}{2} \, \big( \del \D^\m + g^{\m\sb}\del g_{\sb\r} ~ \D^\r \big) ~= - \nabla^{{\rm Ch/H}\, \mb}\Delta_{\a\mb}{}^\n ~.
\eeq

We then find that
$$
\Db^\dag \ccY^{(1)}_\a \= 0~,
$$
 is precisely the D-term equations \eqref{eq:adjoint1}-\eqref{eq:adjoint3}.

Hence, in holomorphic gauge with spurious degrees of freedom we find that \\
$$
\Db \ccY_\a^{(1)} \= 0  \quad \Longleftrightarrow\quad \text{F-terms}~, \qquad\qquad \Db^\dag \ccY_\a^{(1)} \= 0 \quad \Longleftrightarrow \quad \text{D-terms}~.
$$\\
This is despite the fact the individual deformations such as $\fD_\a \A$ or $\D_\a{}^\m$ are not harmonic representatives as demonstrated explicitly in \eqref{eq:HodgeHet1}, \eqref{eq:HodgeHeterotic2a}--\eqref{eq:HodgeHeterotic2c}. Instead,  if we work $(0,1)$--forms valued in $Q$ then all we need to do is study harmonic representatives of this $\Db$--operator. However, we are not yet at the physical case yet: we still need to eliminate the spurious degrees of freedom.

\section{Eliminating the spurious degrees of freedom}
\label{s:nonspur}
The space of deformations of $\Db$ on the bundle $Q$ is not the moduli space of string theory. String scattering amplitudes and/or supersymmetry tell us that $\fD_\a \Th$ is not an independent degree of freedom, but determined in terms of the remaining fields. In \cite{Anguelova:2010ed,Candelas:2018lib}  this is calculated to lowest order in $\ap$:
\beq\label{eq:fdTh}
 \fD_{\a}\Th{}_\mb{}^\n{}_{\s} \= \nabla_\s \, \D_{\a\mb}{}^\n + \half \, \nabla^{\n} \, \ccZ_{\a\,\s\mb}  ~,
\eeq
where 
\beq\notag
  \nabla_\s\,\D_{\a\mb}{}^\n\=\del_\s\,\D_{\a\mb}{}^\n+\G_\s{}^\n{}_\l\,\D_{\a\mb}{}^\l~,\quad \text{and}\quad
  \nabla_{\mb}\,\ccZ_{\a\,\s\nb}\=\del_{\mb}\,\ccZ_{\a\,\s\nb}-\G_{\mb}{}^\lb{}_\nb \,\ccZ_{\a\,\s\lb}~.
\eeq
The $\G$ are the Levi--Civita symbols, which because this result is derived for solutions which have a smooth limit $\ap\to0$ with $H\to 0$, are the same as the Chern connection. The dilaton is constant to $\ap^3$ \cite{Anguelova:2010ed} by a choice of gauge fixing; see also appendix C  of \citeSG for the calculation in the notation of this paper.

\subsection{Constructing the bundle $\cQ$}
We now turn to constructing an extension bundle $\cQ$ whose sections describe the physical degrees of freedom. 
Recall that to first order in $\ap$, the first order deformations satisfy an Atiyah equation $\delb_\th( \fD_\a \th )= \D_\a{}^\m R_\m$. We  check \eqref{eq:spinmoduli} satisfies this explicitly. First, note
\beq\notag
\begin{split}
 \nabla_\rb \nabla_\n \D_\a{}^\m &= \nabla_\rb \left(\del_\n \D_\a{}^\m + \G_\n{}^\m{}_\l\,  \D_{\a}{}^\l  \ \right)~,\\
[\nabla_\rb, \nabla_\n] \D_{\a\sb\,}{}^\m &=   (\del_\rb\G_\n{}^\m{}_\l)\D_{\a\sb\,}{}^\l + (\del_\n \G_\rb{}^\lb{}_\sb)\D_{\a\lb\,}{}^\m   \= R^\m{}_{\l\rb\n} \D_{\a\sb\,}{}^\l - R^\lb{}_{\sb\rb\n} \D_{\a\lb\,}{}^\m ~,
\end{split}
\eeq
where $R^\m{}_{\l\rb\n}$ is the Riemann curvature tensor on a complex manifold.
Second,
\beq\label{eq:AtiyahProof}
\begin{split}
 \delb (\fD_\a \Th)^\m{}_\n &= \dd x^\rb\, \nabla_\rb \left( \nabla_\n \D_\a{}^\m + \ii \nabla^\m \fD_\a \o_\n^{(0,1)} \right)\\ 
 &= \dd x^\rb [ \nabla_\rb , \nabla_\n] \D_\a{}^\m + \half \dd x^\rb [\nabla_\rb, \nabla^\m] \ccZ_{\a\n}^{(0,1)} + \half \nabla^\m (\delb \ccZ_{\a\n}^{(0,1)})\\
 &= \dd x^{\rb\sb} R^\m{}_{\l\rb\n} \D_{\a\,\sb}{}^\l + \dd x^{\rb\sb} R^\lb{}_{\sb\n\rb} \D_{\a\lb\,}{}^\m \\
 &= \D_{\a\,}{}^\l R^\m{}_{\n\l}{}^{(0,1)}~,
\end{split}
 \eeq
 where $R^\lb{}_{\sb\n\rb} = R^\lb{}_{\rb\n\sb}$ is used in the penultimate line. We use that $R^{(0,2)} = 0$ and $\delb \ccZ_\a = \cO(\ap)$. In fact because $\th$ is an instanton (to first order in $\ap$), $\fD_\a \Th$ obeys an Atiyah equation and so we conclude that this result actually holds up to and including first order in $\ap$. Hence, if there are any $\ap$ corrections to \eqref{eq:spinmoduli} then we know that in \eqref{eq:AtiyahProof} that they must cancel. 

A corollary of \eqref{eq:AtiyahProof} is that the expression \eqref{eq:fdTh} is a solution of $\Db \ccY^{(1)}_\a = 0$ and so bona-fide element of the cohomology $H_\Db^1(\ccX, Q)$. This is serves as another consistency check and is part of our intuition that the true moduli space is some subspace of $H_\Db^1(\ccX, Q)$.

Using  the symmetry  $\fD_{\a}\Th{}_\mb{}^\nb{}_{\sb} \=\! - g^{\nb\l} \, \fD_\a\Th_{\mb}{}^\r{}_{\l} \, g_{\r\sb}$ we can  write
\beq\label{eq:LCSevaluated}
\frac{\ap}{2} \tr \big( \fD_\a \Th \, R\big) \= \ap \, \big( \nabla_\n \D_\a{}^\m + \half \, \nabla^\m \, \ccZ_{\a\,\n}^{(0,1)} \big) R^\n{}_\m ~,
\eeq
where $R^\n{}_\m = R_{\r\sb}{}^\n{}_\m \, \dd x^{\r\sb}$ is the Riemann tensor and we have used $R_{\mb\n} = - R_{\n\mb}$. 

Now, using \eqref{eq:AtiyahProof} with the Bianchi identity for $R$,  \eqref{eq:BianchiR} we find 
\beq\label{eq:delbClosed}
\delb\,\left( \tr \fD_\a \Th^{(0,1)} R\right) \= \D_\a{}^\m\, \tr  R_\m \w R~  + \cO(\ap^2)~.
\eeq

Using \eqref{eq:LCSevaluated} we  see that a consequence of \eqref{eq:delbClosed} is that 
$$
\delb \Big(( \nabla^\m \, \ccZ_{\a\,\n}^{(0,1)} ) R^\n{}_\m\Big) \= 0~. 
$$
and so it should be $\delb$--exact. Indeed, this is the case and we now derive an explicit expression. Start by recalling that $R^\n{}_\m$ obeys a hermitian Yang-Mills equation $\nabla^\m R^\n{}_{\m\r\sb} = 0$ to first order in $\ap$ and so 
$$
( \nabla^\m \, \ccZ_{\a\,\n}^{(0,1)} ) R^\n{}_\m \=  \nabla^\m \, (\ccZ_{\a\,\n}^{(0,1)} R^\n{}_\m)~.
$$
As the Levi--Civita symbols are pure to this order in $\ap$ we find the term in parenthesis can be written as
$$
\ccZ_{\a\,\n}^{(0,1)} R^\n{}_\m \=\!-\dd x^{\r} [\nabla_\r,\delb] \ccZ_{\a\,\m}^{(0,1)} ~,
$$
so that
$$
( \nabla^\m \, \ccZ_{\a\,\n}^{(0,1)} ) R^\n{}_\m \=\! - \dd x^{\r} \nabla^\m \left( [\nabla_\r,\delb] \ccZ_{\a\,\m}^{(0,1)}  \right) ~.
$$
Using that $\delb (\ccZ_{\a\,\m}^{(0,1)}) = \cO(\ap)$ and that $R^{(0,2)} = 0$ we find 
$$
( \nabla^\m \, \ccZ_{\a\,\n}^{(0,1)} ) R^\n{}_\m \=\! -\delb \left( \dd x^\r \nabla^\m \nabla_\r  \ccZ_{\a\,\m}^{(0,1)}\right) ~.
$$
Note that $\nabla_\r  \ccZ_{\a\,\m}^{(0,1)} = \del_\r  \ccZ_{\a\,\m}^{(0,1)} - \G_\r{}^\l{}_\m  \ccZ_{\a\,\l}^{(0,1)}$ and so does not become $\del$. 

Using these results \eqref{eq:ccZeomgauge} becomes:
\beq\label{eq:moduli3}
\delb \Big( \ccZ_\a^{(1,1)}- \frac{\ap}{2} \dd x^\r \left(\nabla^\m \nabla_\r  \ccZ_{\a\,\m}^{(0,1)}\right)  \Big) -2\ii \,   \D_\a{}^\m (\del\o)_\m + \ap  R^\n{}_\m\nabla_\n \big(\D_\a{}^\m\big)  -\frac{\ap}{2} \tr \big( \fD_\a \A\, F \big) \= 0 ~.
 \eeq
It is natural to perform a field redefinition
\beq\label{eq:Zrdef}
\wt \ccZ_\a^{(1,1)}\= \ccZ_\a^{(1,1)}- \frac{\ap}{2} \dd x^\r \nabla^\m \nabla_\r  \ccZ_{\a\,\m}^{(0,1)}~,
\eeq
which is well--defined in $\ap$--perturbation theory, and in particular, invertible. It is not overly surprising that the complexified hermitian form needs to be corrected in $\ap$. The field redefinition will propagate into other relevant equations, such as deformations of the HYM and balanced equations and the moduli space metric. But it is the natural thing to do to realise a holomorphic structure on the extension bundle.

There is a cohomology that captures these field deformations. To describe it, we need to generalise some of the above results above to $(0,p)$--forms. First, we ansatz that   \eqref{eq:spinmoduli} generalises to
\beq\label{eq:dthp}
\d_\a \th^{(p)\,\n}{}_\s \=  \nabla_\s \, \D^{(p)}_{\a}{}^\n + \half \, \nabla^{\n} \, \ccZ_{\a\,\s}^{(p)} ~.
\eeq
Second we need the generalisation of the $\Db$--operator for the spurious case, written in components in \eqref{eq:DbCpt}. The only non--trivial part is to evaluate $\tr \d_\a \th^{(p)} R_\n$ using \eqref{eq:dthp} and substitute into the first line of \eqref{eq:DbCpt} giving
$$ 
\wh{\wt \ccZ}^{(p+1)}_\n\= -\delb  {\wt \ccZ}_{\a\,\n}^{(p)} +2\ii \,   \D^{(p)}_\a{}^\m (\del\o)_{\m\n} + \ap   \nabla_\s \big(\D^{(p)}_\a{}^\t\big) R^\s{}_\t{}_{\n\lb} \dd x^\lb  +\frac{\ap}{2} \tr \big( \d_a \A^{(p)}\, F_\n \big)  ~.
$$
A consistency check is that the relation \eqref{eq:dthp} satisfies the second line of \eqref{eq:DbCpt}. We also have the field redefinition 
\beq
\wt \ccZ_{\a\,\n}^{(p)}\= \ccZ_{\a\,\n}^{(p)}+ \frac{\ap}{2} (-1)^p\nabla^\m \nabla_\n  \ccZ_{\a\,\m}^{(p)}~.
\eeq

Now construct the extension bundle. The  extension $Q_1$ defined in \eqref{eq:Q1def} is unchanged.    $Q_2$ is now not necessary. Consider a short exact sequence
 \beq\label{eq:newQ}
\xymatrix{
0 \ar@{>}[r] &\ccT^{\raisebox{1pt}{$*$}}_\ccX{}^{(1,0)}  \ar@{>}[r]^{i}  &  \cQ \ar@{>}[r]^{\pi}& Q_1   \ar@{>}[r] & 0}~,
\eeq
where $\ccT^{\raisebox{1pt}{$*$}}_\ccX{}^{(1,0)}$ are holomorphic co-vectors (forms). 

$p$-forms valued in $\cQ$ are denoted as in the previous subsection 
$$
 \ccY^{(p)}_{\a}\= 
\begin{pmatrix}
 \wt\ccZ^{(p)}_{\a\,\n}  \\
\d_\a \A^{(p)} \\
 \D^{(p)}_\a{}
\end{pmatrix}~.
 $$
We have abused notation in using the same letter $\ccY_\a^{(p)}$ as for the spurious case, but it should be clear from context what we mean.  

Define the operator $\cDb$ on sections valued in this bundle as
$$
\cDb \= 
\begin{pmatrix}
- \delb   & \cH^{new} \\
0 & \Db_1
\end{pmatrix}~,
$$
where the map $\cH^{new}:\O^{(0,p)}(\ccX, Q_1) \to \O^{(0,p+1)}(\ccX,\ccT_\ccX^{*\,(1,0)})$ is
\beq
\cH_\n^{new}( \d_\a \A^{(p)},\Delta^{(p)}_\a) \= 2\ii  \Delta^{(p)}_\a{}^\mu ( \del \o)_{\mu\n}\, - \ap \nabla_\r \big(\D^{(p)}_\a{}^\l\big) R^\r{}_{\l\,\n} 
+  \frac{\ap}{2\phantom{`}}\tr (\d_\a \A^{(p)} F_\n)
~,
\notag\eeq
and where $R^\r{}_{\l\,\n} = R^\r{}_{\l\,\n\sb}\dd x^\sb$. 
The action of $\cDb$ is
\beq\label{eq:Dbaction}
\cDb \begin{pmatrix}
\wt \ccZ_{\a\,\n}^{(p)} \\
 \d_\a \A^{(p)} \\
 \D_\a^{(p)}
\end{pmatrix} \=  \begin{pmatrix}
-\delb \wt\ccZ_{\a\,\n}^{(p)}  + 2\ii  \Delta^{(p)}_\a{}^\mu ( \del \o)_{\mu\n}\, - \ap  (\nabla_\r \D^{(p)}_\a{}^\l) R^\r{}_{\l\,\n}
+  \frac{\ap}{2\phantom{`}}\tr (\d_\a \A^{(p)} F_\n)  \\
\delb_\A (\d_\a \A^{(p)} ) + F_\m \D_\a^{(p)\,\m} \\
\delb \D^{(p)}_\a
\end{pmatrix} 
~.
\eeq
This operator satisfies  $\cDb^2 = 0$ after using the Bianchi identity \eqref{eq:Bianchi2}.\footnote{The operator $\cDb$ squares to {\it zero on the nose} if one modifies the Hull-Strominger system to use the curvature of the zeroth order Ricci-flat Calabi-Yau metric in the $\tr\,R^2$ term in the heterotic Bianchi identity. Otherwise, this only holds modulo $\ap^2$ corrections.} Furthermore,  $\cDb \ccY_\a^{(1)} = 0$ corresponds to precisely the F-term equations  \eqref{eq:modulieqnHolGauge}--\eqref{eq:ccZeomgauge} together with \eqref{eq:fdTh}. It should be noted that in contrast to the spurious case of section \ref{s:spur}, this operator is not a connection in that for a function $f$ and section $\ccY^{(p)}_{\a}$ of $ \cQ$ the Lebniz rule is not satisfied:
\begin{equation}
\cDb(f\,\ccY^{(p)}_{\a})\neq \delb f\wedge\ccY^{(p)}_{\a}+f\cDb\ccY^{(p)}_{\a}\:.
\end{equation}
This is due to the appearance of a holomorphic derivative in the new extension map $\cH_\n^{new}$.\footnote{We are grateful to Mario Garcia Fernandez for pointing this out.}

\subsection{The D-terms and the adjoint operator $\overline{\mathcal{D}}^\dag$}
The moduli space metric, after including \eqref{eq:spinmoduli}, is \citeCOV
\beq\begin{split}\label{eq:modulimetricnospurious}
 g\#_{\a\bb} &\= \frac{1}{V}\int_X \Big( \D_\a{}^\m\star\D_{\bb}{}^{\nb}\,g_{\m\nb} + \frac{1}{4} \ccZ_\a^{(1,1)}\star\ccZb_\bb^{(1,1)} + \frac{\ap}{4}\tr{ ( \fD_\a A \star \fD_{\bb}A ) } \Big)\\[0.1cm]
 &\hspace{3cm}~+~ \frac{\ap}{2V}\int_X \vol ~ \Big( \D_{\a\mb\nb}\,\D_{\bb\r\s} + \frac{1}{4} \ccZ_{\a\,\r\mb}\,\ccZb_{\bb\,\s\nb} \Big) \, R^{\r\mb\s\nb} +  \cO(\ap^2) ~,
\end{split}\raisetag{2cm}\eeq
where we have used holomorphic gauge \eqref{eq:finalGauge} in writing $\ccZ_\a^{(1,1)} = 2\ii \fD_{\a}\o^{(1,1)}$. We need to write down $g\#_{\a\bb}$ in the new variables $\wt \ccZ_\a^{(1,1)}$ via the field redefinition \eqref{eq:Zrdef}. First note that\\
$$
\int_X \ccZ_\a^{(1,1)}\star \ccZb_\bb^{(1,1)} \= \int_X \left\{ \wt \ccZ_\a^{(1,1)} \star  \ccZtb_\bb^{(1,1)} - \frac{\ap}{2}  \big(\nabla^\m\nabla_\r \wt \ccZ_{\a\,\m}^{(0,1)} \big) \dd x^\r \star  \ccZtb_\bb^{(1,1)}
- \frac{\ap}{2}\wt \ccZ_\a^{(1,1)}  \star \big(\nabla^\mb\nabla_\rb  \ccZtb_{\bb\,\mb}^{(1,0)}\dd x^\rb \big) \right\}~.\\[5pt]
$$
 The second term is
\beq\notag
\begin{split}
 \int_X \big(\nabla^\m\nabla_\r \ccZt_{\a\,\m}^{(0,1)}\dd x^\r \big) \star \ccZtb_\bb^{(1,1)}& {\=}\!-  \int_X\vol \big(\nabla^\m\nabla_\r \ccZt_{\a\,\m\lb} \big) \ccZtb_{\bb}{}^{\, \r \lb}  \\
 &{\=}  \int_X \big( \nabla^\m\ccZt^{(0,1)}_{\a\,\m} \big) \star \big( \nabla^\tb\ccZtb^{(1,0)}_{\bb\,\s} \big)  {-}  \int_X\vol \Big([\nabla^\m,\nabla_\r ] \ccZt_{\a\,\m\lb} \Big) \ccZtb_{\bb}{}^{\, \r \lb}~.
\end{split}
\eeq
The commutator involves both the Riemann curvature and Ricci tensor $\Ric^{\d\tb} =\! - R^\d{}_\m{}^{\m\tb}$:
$$
[\nabla^\m,\nabla^\tb ] \ccZt_{\a\,\m\lb} \= - R^\d{}_\m{}^{\m\tb} \ccZt_{\a\,\d\lb} - R^\db{}_\lb{}^{\m\tb} \ccZt_{\a\,\m\db} \= \Ric^{\d\tb} \ccZt_{\a\,\d\lb} - R^\db{}_\lb{}^{\m\tb} \ccZt_{\a\,\m\db}~.
$$
The Ricci tensor vanishes to this order in $\ap$. So 
\beq\notag
\begin{split}
\int_X \big(\nabla^\m\nabla_\r \ccZt_{\a\,\m}^{(0,1)}\dd x^\r \big) \star \ccZtb_\bb^{(1,1)}  &\= \int_X \big( \nabla^\m\ccZt^{(0,1)}_{\a\,\m} \big)\star \big( \nabla^\tb\ccZtb_{\bb\,\tb}^{(1,0)} \big) 
   + \int_X  \vol R^{\m\db\s\tb} \ccZt_{\a\,\m\db}
  \ccZtb_{\bb\, \s\tb}~.
\end{split}
\eeq
The third term follows in the same way. 
Hence, the moduli space metric in these variables is
\beq\label{eq:newmodulimetric}
\begin{split}
 g\#_{\a\bb} &\= \frac{1}{V}\int_X \Big( \D_\a{}^\m\star\D_{\bb}{}^{\nb}\,g_{\m\nb} + \frac{1}{4} \ccZt_\a^{(1,1)}\star\ccZtb_\bb^{(1,1)} + \frac{\ap}{4}\tr{ ( \fD_\a A \star \fD_{\bb}A ) } \Big)\\[0.1cm]
 &-\frac{\ap}{4V} \int_X \big( \nabla^\m\ccZt^{(0,1)}_{\a\,\m} \big)\star \big( \nabla^\tb\ccZtb_{\bb\,\tb}^{(1,0)} \big)  -  \frac{\ap}{8V} \int_X  \vol R^{\m\db\s\tb} \ccZt_{\a\,\m\db}   \ccZtb_{\bb\, \s\tb} \\
 &\hspace{3cm}~+~ \frac{\ap}{2V}\int_X \vol ~ \Big( \D_{\a\mb\nb}\,\D_{\bb\r\s}  \Big) \, R^{\r\mb\s\nb}  ~.
\end{split}\raisetag{2cm}\eeq
It is straightforward to generalise this calculation to a metric on $(0,p)$--forms:
\beq\label{eq:bigmetricnospu}
\begin{split}
 \langle\ccY^{(p)}_\a , \ccY^{(p)}_\b \rangle &\=  \frac{1}{V}\int_X \left\{ \frac{1}{4} \wt \ccZ_{\a\,\n}^{(p)} \star  \ccZtb_{\bb}^{(p)\,\n} -\frac{\ap}{4}  \big( \nabla^\m\ccZt^{(p)}_{\a\,\m} \big)\star \big( \nabla^\tb\ccZtb_{\bb\,\tb}^{(p)} \big)  
 \right.\\[3pt]
 &  \qquad\qquad + \frac{\ap}{8} (-1)^p R^{\m\db\s\tb} \ccZt^{(p)}_{\a\,\m\db} \star  \ccZtb^{(p)}_{\bb\, \s\tb} - \frac{\ap}{4}\tr{ ( \d_\a \A^{(p)} \star \d_{\bb}\A^{(p)}{}^\dag} )\\[3pt]
&\qquad\qquad\left.  + \D^{(p)}_\a{}^\m\star\D^{(p)}_{\bb}{}^{\nb}\,g_{\m\nb} + \frac{\ap}{2}  (-1)^{p+1}R^{\mb\r}{}_{\n\sb} \D_{\a\,\mb}^{(p)}{}^\n\star \D^{(p)}_{\bb\,\r}{}^\sb \,  \right\}~,
\end{split}
\eeq
where $\D_{\a\,\mb}^{(p)}{}^\nb = \frac{1}{(p-1)!} \D^{(p)}_{\a\,\mb\lb_1\cdots\lb_{p-1}}{}^\nb\dd x^{\lb_1\cdots\lb_{p-1}}$. For $p=0$, the third and last terms are not present. 
We have also used
$$
\frac{\ap}{4}  \tr (\d_\a \th^{(p)} \star \d_\bb \th^{\dag\,(p)}) \= \frac{\ap }{2} (-1)^{p+1} \left\{ \frac{1}{4} (\ccZ_\a^{(p)})_{\m\db} \star (\ccZb_\bb^{(p)})_{\s\tb} R^{\m\db\s\tb}  + (\D_\a{}^{(p)})_\mb{}^\n \star (\D_\bb^{(p)})_\s{}^\tb R^{\mb\s}{}_{\n\tb}\right\}~.
$$

Using the derivative operator in \eqref{eq:Dbaction}  and the metric above, we compute the adjoint following the logic of \sref{s:spuradjoint}. Many of the terms follow through identically. For ease of notation, we now drop the tilde on $\ccZ_\a^{(p)}$. The adjoint is 
\beq\notag
\begin{split}
 \langle\cDb \ccY^{(p)}_\a , \ccY^{(p+1)}_\b \rangle &\=  \frac{1}{V}\int_X \left\{ \frac{1}{4} \Big[ - \delb \ccZ_{\a\,\n}^{(p)} +\frac{\ap}{2} \tr \big( \d_\a \A^{(p)} \, F_\n\big) +2\ii \Delta^{(p)}_\a{}^\mu ( \del \o)_{\mu\n}\, - \ap  \nabla_\s \D^{(p)}_\a{}^\t R_{\n\lb}{}^\s{}_\t \dd x^\lb  \Big] \star  \ccZb_\bb^{(p+1)\,\n}\right.\\
 &\qquad +\frac{\ap}{4} \delb \big( \nabla^\s \ccZ^{(p)}_{\a\,\s} \big)\star \big( \nabla^\tb\ccZb_{\bb\,\tb}^{(p+1)} \big)  
  +  \frac{\ap}{8}(-1)^p   R^{\s\lb\t\nb} \big(-\delb \ccZ^{(p)}_{\a} \big)_{\s\lb} \star  \ccZb^{(p)}_{\bb\, \t\nb} \\[3pt]
  &\qquad - \frac{\ap}{4}\tr{\Big[ \Big( \delb_\A \d_\a \A^{(p)} + F_\m \D^{(p)}_\a){}^\m \Big) \star \d_{\bb}\A^\dag{}^{(p+1)}{}} \Big]\\[3pt]
&\qquad\qquad\left.  + (\delb\D^{(p)}_\a){}^\m\star\D^{(p+1)}_{\bb}{}^{\nb}\,g_{\m\nb} + \frac{\ap}{2} (-1)^{p+1} R^{\mb\r}{}_{\n\sb} (\delb \D^{(p)}_{\a})_\mb{}^\n\star (\D^{(p+1)}_{\bb})_{\r}{}^\sb \,  \right\}\\[3pt]
&\= \frac{1}{V} \int_\ccX \left\{ \frac{1}{4} \ccZ_{\a\,\n}^{(p)} \star \Big[-\del^\dag \ccZb_\bb^{(p+1)\,\n} + \frac{\ap}{2} (-1)^p R^{\t\sb\n\lb} \nabla_\lb \ccZb^{(p+1)}_{\bb\,\t\sb} \Big]
\right.\\[5pt]
& \phantom{=\frac{1}{V} \int_\ccX \Big\{ ~} -\frac{\ap}{4} \nabla^\m \ccZ_{\a\m}^{(p)} \star \nabla^\tb (-\del^\dag \ccZb_{\bb}^{(p+1)})_\tb +\frac{\ap}{8} (-1)^p R^{\t\nb\s\db} \ccZ^{(p)}_{\a\,\s\db} \star \big( - \del^\dag \ccZ^{(p+1)}_{\bb\,\t\nb} \big) \\[3pt]
&\quad - \frac{\ap}{4} \tr \left( \d_\a \A^{(p)} \star \left( \del^\dag_{\A^\dag} \d_\bb \A^{\dag\,(p+1)} + \frac{1}{2}(-1)^{p+1}  F_{\m\nb} \ccZ_\bb{}^{\m\nb\,(p+1)}   \right) \right)
\\[3pt]
&\!\!\!\!\!\!\!\!\!\!\!\!+ \D^{(p)}_\a{}^\m \star \left[ (-1)^p \frac{\ii}{2} (\del\o)^{\r\nb\lb} \ccZb_{\bb\,\r\nb}^{(p+1)}  - \frac{\ap}{4}  \tr F^{\lb\n}\, \d_\bb \A_\n^\dag{}^{(p+1)}  + \del^\dag \D_{\bb}^{(p+1)}{}^\lb - \frac{\ap}{4}(-1)^p R^{\lb\n\s\tb} \nabla_\n \ccZb_{\bb\,\s\tb}^{(p+1)}\right.\\[3pt]
&\qquad\qquad\qquad\left. \left. + \frac{\ap}{2}(-1)^p  R^{\lb\r}{}_{\m\sb} \nabla_\lb \D^{(p+1)}_{\bb\,\r}{}^\sb  \right] g_{\m\lb} + \frac{\ap}{2} (-1)^{p+1} \D^{(p)}_{\a\,\nb} \star (\del^\dag \D^{(p+1)}_\bb )_\r{}^\sb R_{\m\sb}{}^{\r\nb} \right\}\\[5pt]
&\=  \langle  \ccY_\a^{(p)}, \Db^\dag\ccY_\b^{(p+1)} \rangle~.
\end{split}
\eeq
We compare with the metric \eqref{eq:bigmetricnospu} and work to first order in $\ap$ to deduce the action of the adjoint operator on sections
\beq\label{eq:Dadj}
 \cDb^\dag  \ccY_\a^{(p+1)} \= 
 \begin{pmatrix}
\wh \ccZ_\a^{(p)} \\
\wh \d_\a \A^{(p)} \\
\wh \D_\a^{(p)}{}^\m
\end{pmatrix}~,
\eeq
where
\beq
\begin{split}
 \wh \ccZ_\a^{(p)} &\= -\delb^\dag \ccZ_\a^{(p+1)\,\nb} - \frac{\ap}{2} (-1)^p R^{\t\sb\l\nb} \nabla_\l \ccZ^{(p+1)}_{\a\,\t\sb}~, \\[3pt]
 \wh \d_\a \A^{(p)}&\=  \delb_\A^\dag \d_\a \A^{(p+1)} + \half (-1)^p F^{\r\nb} \ccZ^{(p+1)}_{\a\,\r\nb}~,\\[3pt]
\wh \D_\a^{(p)}{}^\m &\= (-1)^{p+1} \frac{\ii}{2} (\delb\o)^{\m\r\lb} \ccZ^{(p+1)}_{\a\,\r\lb} + \frac{\ap}{4} \tr (F^{\m\rb}\, \d_\a \A^{(p+1)}_\rb) - \nabla^{{\rm Ch/H}\, \nb}\Delta^{(p)}_{\a\nb}{}^\m \\
&\qquad\qquad +  \frac{\ap}{2}  (-1)^p R^\m{}_{\r}{}^{\s\tb} \nabla_\s\D^{(p)}_{\a\,\tb}{}^\r   - \frac{\ap}{4}(-1)^p R^{\m\rb\s\tb} \nabla_\rb \ccZ^{(p)}_\a{}_{\s\tb}{}~.\\[3pt]
\end{split}
\eeq	
That is, the first line is a component of the balanced equation \eqref{eq:coclosedZ} after the redefinition \eqref{eq:Zrdef}; the second line is the vanishing of the HYM equation \eqref{eq:adjoint2}; the third line is the other component of the balanced equation \eqref{eq:adjoint1}, where we use \eqref{eq:spinmoduli}. 

The case $\cDb^\dag \ccY_\a^{(1)} = 0$ amounts to 
 precisely the D-terms. 
 
\section{Conclusion and Outlook}

So we have shown that physical field deformations in holomorphic gauge are given by harmonic representatives of $\cDb$ operator on the new $\cQ$ bundle with the spurious degrees of freedom eliminated. These are harmonic with respect to the moduli space metric in \citeSG. We see\\
$$
\cDb \ccY_\a^{(1)} \= 0  \quad \Longleftrightarrow\quad \text{F-terms}~, \qquad\qquad \cDb^\dag \ccY_\a^{(1)} \= 0 \quad \Longleftrightarrow \quad \text{D-terms}~.
$$\\
Along the way we also demonstrated that the non--physical calculation, in which one has additional spurious degrees of freedom, are captured by a $\Db$--operator with an analogous relation between the F-terms and D-terms and $\Db$--closure and $\Db$--coclosure. That means that with respect to the $\Db$--operator we could equally well call holomorphic gauge, harmonic gauge. 

It would be interesting to reconcile these results with the complexified gauge transformations and moment maps in a GIT-type analysis of these solutions studied in \cite{Garcia-Fernandez:2020awc, Ashmore:2019rkx}. There are also examples of mathematical interest, such as that related to the Hopf surface which do not have an obvious valid $\ap$--expansion, yet appear to be governed by a $(0,2)$ superconformal algebra \cite{Alvarez-Consul:2020hbl}. Nonetheless, one could still try to construct the adjoint operator $\cDb^\dag$ and compare with the space of deformations obtained. It would also be interesting to see to what extent the structure discovered in this paper extends to next order in $\ap$. We expect the generalised geometry approach of \cite{Garcia-Fernandez:2015hja, Garcia-Fernandez:2020awc, Garcia-Fernandez:2018emx, Garcia-Fernandez:2018ypt, Ashmore:2019rkx} to be relevant for these scenarios. 

One could also ask about higher orders in deformation theory. Presumably at second and third order in deformation theory there are relations with the $L_3$ algebra described in \cite{Ashmore:2018ybe}. It is known in the study of CY manifolds that special geometry involves relations between the moduli and matter sectors. It would be interesting to extend these results to the charged matter sector utilising the matter metric \citeMatter. The results here are reminiscent of the study of the $tt^*$ equations of the special geometry of CY manifolds \cite{Cecotti:1991me}, in which one studies a fiber bundle whose base manifold is the moduli space and fibres are harmonic elements of the cohomologies $H^{(2,1)} \oplus H^{(1,1)}$. Until now, we haven't been able to demonstrate a cohomology with harmonic representative capturing the deformations of the full heterotic theory. This might be a new door worth opening.

\section*{Acknowledgements}
We would like to thank A. Ashmore, A. Coimbra, X. de la Ossa, M. Garcia Fernandez, M. Larfors, M. Magill, R. Sisca, C. Strickland-Constable, G Tartaglino--Mazzucchelli, D. Tennyson for useful discussions.

\bigskip
\appendix
\section{Some useful relations}
\subsection{Curvatures and Bianchi identity}
The Bianchi identity for a curvature $R$ evaluated with a connection $\Th$ is
\beq\label{eq:BianchiR}
\dd R^m{}_n + \Th^m{}_p R^p{}_n -   R^m{}_p \Th^p{}_n \= 0 ~, \quad \Leftrightarrow \quad \dd_\Th R \= 0~.
\eeq

 For now, we take $R$  to be the Riemann curvature tensor. It has symmetry properties
$$
R_{mnpq} = - R_{mnqp} = - R_{nmpq} = R_{pqmn}~.
$$

We have have need for commutators of derivatives to lowest order in $\ap$ e.g.
$$
[\nabla^\m,\nabla^\tb ] \ccZ_{\a\,\m\lb} \= - R^\d{}_\m{}^{\m\tb} \ccZ_{\a\,\d\lb} - R^\db{}_\lb{}^{\m\tb} \ccZ_{\a\,\m\db} \= \Ric^{\d\tb} \ccZ_{\a\,\d\lb} - R^\db{}_\lb{}^{\m\tb} \ccZ_{\a\,\m\db}~.
$$
It involves both the Riemann curvature and Ricci tensor $\Ric^{\d\tb} =\! - R^\d{}_\m{}^{\m\tb}$:
The Ricci tensor vanishes to this order in $\ap$.

It is useful to note that
$$
2\ii \delb(\delb \o)_{\m\n} \= 2\ii (\delb\del\o)_{\m\n} \= - \frac{\ap}{4} \tr (F^2)_{\m\n} = \frac{\ap}{2} \tr F_\m F_\n~,
$$
where we omit the $\tr R^2$ term as it follows in the obvious way with a minus sign. 

\subsection{Hodge dual relations}
It is useful to recount from the appendix of \citeSG adjoint differential operators and some Hodge dual relations for forms on $X$.

The Hodge $\star$ operator acts on type
\beq\notag
 \star : \O^{(p,q)}(X) \to \O^{(N-q,N-p)}(X) ~.
\eeq

Given $k$-forms $\eta$, $\x$ and a metric $\dd s^2 = g_{mn} \dd x^m \otimes \dd x^n$ on $X$, the Hodge dual defines an inner product
\beq\notag
 (~\cdot~,~\cdot~) ~~: ~~ \O^k(X) \times \O^k(X) \to \mathbb{R} ~,
\eeq
with
\beq\notag
 (\eta,\xi)     \=  \frac{1}{V\, k!} \int_X \vol \, \eta^{m_1 ... m_k} \, \xi_{m_1 ... m_k}  ~.
\eeq
Now consider two forms $\eta_k$ and $\xi_l$, where the subscript denotes their degree and $k \leq l$,. Contraction is 
\beq\notag
 \lrcorner : \O^k(X) \times \O^l(X) \to \O^{l-k}(X) ~,
\eeq
and acts as follows
\beq\notag
 \eta_k \, \lrcorner \, \xi_l  \=  \frac{1}{k!(l-k)!} \, \eta^{m_1 ... m_k} \, \xi_{m_1 ... m_k \, n_1 ... n_{l-k}} \, \dd x^{n_1 ... n_{l-k}}  \=  \frac{1}{k!} \,\eta^{m_1 ... m_k} \, \xi_{m_1 ... m_k} ~.
\eeq
An interesting feature of this operator is that it is the adjoint of the wedge product
\beq\label{contradjwedg}
 (\s_{l-k} \, \lrcorner \, \xi_l , \eta_k)  \=  (\xi_l , \s_{l-k} \w \eta_k) ~.
\eeq
Recall, the de Rahm operator $\dd$ can be written
\beq\notag
 \dd  \=  \dd x^m \, \nabla^{\LC}_m ~.
\eeq
Using  \eqref{contradjwedg} and  integration by parts
\beq\notag
 ( \dd \eta , \xi )  \=  ( \dd x^m \, \nabla^{\LC}_m \eta , \xi )  \=  ( \nabla^{\LC}_m \eta , \xi^m )  \=  ( \eta , - \nabla^{\LC}_m \xi^m ) ~.
\eeq
It follows that
\beq\notag
 \dd^\dag \xi_k  \=\!  - \nabla^{\LC}_m \xi^m  \=\!  - \frac{1}{(k-1)!} \, \nabla^{\LC}_n \xi^n{}_{m_1 ... m_{k-1}} \, \dd x^{m_1 ... m_{k-1}} ~.
\eeq

The de Rham differential splits into the sum of Dolbeault operators $\dd = \del + \delb$. Analogously, the codifferential also splits $\dd^\dag = \del^\dag + \delb^\dag$ where
\beq\label{eq:deladjoint}
\begin{split}
 &\del^\dag : \O^{(p,q)}(X) \to \O^{(p-1,q)}(X) \qquad , \qquad \del^\dag  \=\!  - \star \delb \, \star ~,\\[0.1cm]
 &\delb^\dag : \O^{(p,q)}(X) \to \O^{(p,q-1)}(X) \qquad , \qquad \delb^\dag  \=\!  - \star \del \, \star ~.
\end{split}\eeq

\vskip0.5cm
 One-forms, type $(1,0)$:
\beq\label{eq:Hodge1form}
 \star\, \eta^{(1,0)} \=\! -\ii \, \eta^{(1,0)} ~ \frac{\o^2}{2} ~.
\eeq
 Two-forms, types $(2,0)$ and $(1,1)$:
\beq\begin{split}\label{eq:Hodge2form}
 &\star\, \eta^{(2,0)}  \=  \eta^{(2,0)} \, \o ~,\\[0.1cm]
 &\star\, \eta^{(1,1)}  \=\! -\ii \, \eta_{\m}{}^\m ~ \frac{\o^2}{2} - \eta^{(1,1)} \, \o \= (\o \, \lrcorner \, \eta^{(1,1)}) ~ \frac{\o^2}{2} - \eta^{(1,1)} \, \o ~,
\end{split}\eeq
 Three-forms, types $(3,0)$ and $(2,1)$:
\beq\begin{split}\label{eq:Hodge3form}
 &\star\, \eta^{(3,0)}  \=\!  - \ii \, \eta^{(3,0)} ~,\\[0.1cm]
 &\star\, \eta^{(2,1)}  \= \ii \, \eta^{(2,1)} - \eta_{\m}{}^\m{}\,^{(1,0)} \, \o \= \ii \, \eta^{(2,1)} - \ii \, (\o \, \lrcorner \, \eta^{(2,1)}) \, \o ~,
\end{split}\eeq
 Type $(2,3)$:
\beq\label{eq:Hodge23}
 \star \eta^{(2,3)} \= \frac{\ii}{2} ~ \eta_{\m\n}{}^{\m\n}{}\,^{(0,1)} \= \frac{\ii}{2} ~ \o \, \lrcorner \, (\o \, \lrcorner \, \eta^{(2,3)}) ~.
\eeq

\section{A supergravity no-go theorem}
\label{app:NoGo}
No--go theorems are well-known in the context of type II and M-theory flux vacua. Classical supergravity equations of motion, when integrated over a compact manifolds, force all non--trivial fluxes to vanish. The evasion of these theorems comes in the form of $\ap$--corrections to supergravity,  interpreted as the contribution of orientifold planes, objects with negative stress--energy tensor. Here we briefly recount an analogous result in heterotic string theory; this demonstrates that if one wants solutions of supergravity then $H=\cO(\ap)$.

In this section only we work to $\ap^3$, as it is not complicated to include the $\ap^2$ correction. The equations of motion are given by
\begin{equation}\begin{split}
&  \cR - 4(\nabla \Phi)^2 + 4 \nabla^2 \Phi - \half |H|^2 - \frac{\alpha'}{4} \big(\tr |F|^2 - \tr |R(\Th^+)|^2\big) + \cO(\ap^3) \= 0~,\\[3pt]
& \cR_{MN}+ 2 \nabla_M \nabla_N\Phi - \frac{1}{4} H_{MAB} H_N{}^{AB} -
\frac{\alpha'}{4} \Big( \tr F_{MP} F_N{}^P - R_{MPAB}[\Th^+]R_N{}^{PAB}[\Th^+]\Big)  + \cO(\ap^3) \= 0~,\\[6pt]
&\nabla^M(\ee^{-2\Phi} H_{MNP}) + \cO(\ap^3) \= 0~,\\[8pt]
&\ccD^{-\,M} (\ee^{-2\Phi} F_{MN}) + \cO(\ap^3) \= 0~,
\end{split}\raisetag{45pt}\label{EOM}\end{equation}
\vskip5pt
here $M,N=0,\cdots,9$, $\ccD^- = \nabla^- + [A,\cdot]$,  with $\nabla^-$ computed with respect to the $\Th^-$ connection, and $\cR_{MN}$ is the Ricci tensor. We restrict to the internal manifold $X$ directions and by tracing the second equation, subtracting form the first we find
\beq
 2 \nabla^2 \Phi \= -  4(\nabla \Phi)^2 +  |H|^2 + \frac{\alpha'}{4} \big(\tr |F|^2 - \tr |R(\Th^+)|^2\big) ~.\\[3pt]
\eeq
We we want to remain in the regime of perturbative string theory, one in which we might hope to have a worldsheet description via  a $(0,2)$--sigma model or even better a $(0,2)$--CFT, then we require 
$$
\Phi  = \Phi_0 + \alpha'\Phi_1 + \cdots~,
$$
where $\Phi_0$ is a constant scalar on $X$. As we see below this is related to the string coupling constant $g_s$. It is possible there are non--perturbative solutions involving either 5-branes or duality, however these are beyond the scope of what we study. They require tools beyond the Hull--Strominger system to determine if the quantum corrections, in both $\ap$ and $g_s$, are convergent. 

Integrating both sides over $X$ we see
$$
\int  \vol \left\{|H|^2 - 4 (\nabla\Phi)^2\right\} + \frac{\ap}{4} \int_X\vol \big(\tr |F|^2 - \tr |R(\Th^+)|^2\big)   \= 0. 
$$
From the previous paragraph, we see that for $g_s$--perturbative solutions and because  $|H|^2$ is positive definite, that $H=\cO(\ap)$. So if we want to study solutions of string theory with supergravity limits then $H=\cO(\ap)$.  

If we didn't have the contribution of the last term there would be no solution at all -- this is the classical supergravity no-go theorem. We evade this because of the Green--Schwarz anomaly cancellation condition between the classical supergravity contributions and the $\ap$--correction involving the curvature squared terms. In certain backgrounds heterotic theories are dual to type IIB theories, and this contribution can be understood as coming from the contribution of $O7$--planes in type IIB, which carry negative energy density. This explains physically why the contribution from the $\tr R^2$ term in the moduli space metric appears with the opposite sign from all the other terms. That is \citeM, 
$$
 g\#_{\a\bb} \= \frac{1}{V}\int_X \Big( \frac{1}{4} \ccZ_\a^{(1,1)} \star\ccZb_\bb^{(1,1)}   +  \frac{\ap}{4}\tr{ ( \fD_\a A \star \fD_{\bb}\A ) }+ \D_\a{}^\m\star\D_{\bb}{}^{\nb}\,g_{\m\nb}  -\frac{\ap}{4} \tr{ ( \fD_\a \Th \star \fD_{\bb}\Th ) } \Big)~.
$$
This does not mean the moduli space metric has indefinite signature: it is constructed in supergravity and so necessarily $\ap\ll 1$ meaning the last term, no matter its value, cannot change the signature of the metric. Even after substituting for the spurious degrees of freedom, the same logic will apply. 

If one wanted to study solutions where $\ap$ is large, then if we want it to be relevant at all to string theory (along with all its beautiful results such as mirror symmetry, special geometry and sigma models) then we need to include the higher order $\ap$--corrections. It is already known that there is an $\ap^2$ correction to the hermitian sector of the metric \citeAQS. As pointed out in that paper,  there are going to be $\ap^3$ corrections, such as the $\ap^3$ correction to the moduli space metric calculated using mirror symmetry \cite{Candelas:1990rm} at the standard embedding
\beq\label{eq:KahlerPotSG}
\begin{split}
K_{\ap^3} &\= - \log \left( \frac{4}{3} \int_X \o^3 + 8 \ap^3 \zeta(3) \chi(X) \right)\\
 & \approx - \log \left( \frac{4}{3} \int_X \o^3\right) + 8\ap^3 \frac{\zeta(3)\chi(X)}{V} + \cdots~,
\end{split}
\eeq
where $V$ is the volume of $X$, $\zeta$ is the Riemann zeta function and $\chi$ the Euler number. Presumably, if we understood $(0,2)$--mirror symmetry, we could determine all the $\ap$--corrections to the moduli space metric. 

With all deformations turned on, the \K potential for the moduli space metric $g\#_{\a\bb}$ was first shown to be \citeM
\beq\label{eq:KahlerPot}
K \= -\log \left(\frac{4}{3} \int_X \o^3 \right) - \log \left( \ii \int_X \O\, \Ob \right)~.
\eeq
It is remarkable that this also encodes the deformations of the vector bundle despite involving quantities which naively belong to $X$.\footnote{There is additional term coming from the universal axio--dilaton which in the tradition of special geometry we do not write as it is completely decoupled from all the other parameters. We discuss it below.} In the gauge in which the dilaton is constant \citeAQS, the norm $\norm{\O}$ is constant over $X$ and so  the compatibility relation between $\O$ and $\o$ is
$$
\ii \int_X  \O \Ob \= \frac{1}{3!} \int_X \norm{\O}^2 \o^3~.
$$
 Substituting into \eqref{eq:KahlerPot} we see  that
\beq
K \= -\log \left( \int_X \norm{\O} \o^3 \right)~,
\eeq
where we have dropped irrelevant numerical constants.

We now show this is equivalent to the four--dimensional dilaton as written in Einstein frame. Indeed, in Einstein frame of the dimensional reduction of heterotic supergravity to $\IR^{3,1}$, the ten--dimensional dilaton is written\footnote{We again use that modulo an appropriate gauge fixing the dilaton is constant  up to order $\ap^3$ \cite{Anguelova:2010ed}.}
$$
\Phi \= \Phi_0 + \vph(X^e) + \ap^3 \Phi_3(X,x) + \cdots, 
$$
where $X^e = (X^0,\cdots,X^3)$ are the $\IR^{3,1}$ coordinates and $x^m$ are the coordinates of $X$. The first term is the zero mode and so a constant. There is a purely $d=4$ fluctuation $\vph(X)$ and the next non--zero term is $\Phi_3$ which is cubic order in $\ap$. The canonical definition of the  four--dimensional dilaton, even for non-\K manifolds is (see for example \cite{Cassani:2007pq}):
$$
e^{-2\phi_4} \= e^{-2\varphi(X^e)} \frac{1}{ g_s^2 V_0} \int_X\frac{1}{3!} \o^3  + \cO(\ap^3)~,
$$
where $g_s  = e^{\Phi_0}$ is the zero--mode of the dilaton and is the string coupling constant. For the Hull--Strominger system to be a good approximation to string theory, we require $g_s \to 0$. $V_0$ is a reference volume of $X$, measured at any point in moduli space, and appears in order for the dilaton to be dimensionless. 

A consequence of supersymmetry is that 
\beq\label{eq:OmDilaton}
\dd \log \norm{\O} \= \! -2 \dd \Phi + \cO(\ap^2)~,
\eeq
where $\dd$ is the ten-dimensional exterior derivative. We therefore get
$$
e^{-2\phi_4} \=  \frac{1}{ g_s^2V_0 } \int_X\frac{1}{3!} \norm{\O} \o^3 + \cO(\ap^2) ~. 
$$
Supersymmetry here is important. The four--dimensional dilaton appears in a combination with the universal axion -- the $B$--field whose legs span $\IR^{3,1}$: $ S = b + \ii e^{-2\phi_4}$. In the language of $N=1$, $d=4$ supersymmetry  $S$ is a component of a linear multiplet. 

We we see that,  up to irrelevant constants, 
\beq\label{eq:fourdilaton}
2\phi_4  = - \log  \int_X\norm{\O} \o^3 - \log\left( \frac{\ii}{2}  ( \overline{S} - S) \right)+ \cO(\ap^2)~. 
\eeq
The second term we often do not include as the universal axio--dilaton is not coupled to any other fields at this order in $\ap$ and $g_s$  and so its contribution to the moduli space metric is somewhat trivial \citeM.

As promised, we have shown that $ K = 2\phi_4$ and the four-dimensional dilaton is the \K potential  \eqref{eq:KahlerPot}  first written in \citeM. All we demanded here were the physically reasonable assumptions that supergravity is  a good approximation, viz. $\ap,g_s \to 0$.

Supersymmetry does not afford the dilaton functional any protection from $\ap$ corrections. The linear multiplet is a fully--fledged D-term and so will receive corrections at all orders in $\ap$.\footnote{We thank Gabriele Tartaglino--Mazzucchelli for explaining this to us.} That being so, we expect the first correction already at $\ap^2$. Many issues arise. It is not even clear, for example, if the four--dimensional dilaton and the \K potential for the moduli space metric at $\ap^2$ are the same thing. Given the supergravity action and supersymmetry variations are known at this order, determining this correction is an interesting future direction. It would be very interesting to determine the $\ap^3$ corrections as well, though the supersymmetry variations and equations of motion are not completely known at this order. 

Recently, \cite{Garcia-Fernandez:2020awc} took \eqref{eq:fourdilaton} as the \K potential for the moduli space metric to deformations of the first order $\ap$ Hull--Strominger system assuming the complex structure is fixed. They then related this to the Aeppli and Bott--Chern cohomology. In the limit $\ap\to0$ with $H\to0$, in which the Hull--Strominger system accurately describes string backgrounds, it reproduces \citeM as we demonstrated above. Their metric also applies to formal solutions of the Hull-Strominger system one allow for $\ap$ to be large and for non--trivial dilaton with torsion $H=\cO(1)$. However, as discussed above, such solutions are not obviously string solutions because of the non--trivial dilaton behaviour and because the $\ap^2,\ap^3,\cdots$ corrections are not included.  They do not find, for example, the $\ap^3$ correction to the metric predicted from mirror symmetry in the special case of the standard embedding cf. \eqref{eq:KahlerPotSG}. This term will dominate of the lower order contributions in such solutions. Furthermore, interesting pathologies can also arise in the large $\ap$ limit, such as the moduli space metric being degenerate and so not \K as required by $N=1$ $d=4$ supersymmetry. This reinforces the importance of including perturbative $\ap^2,\ap^3,\cdots$--corrections to the Hull--Strominger system --- they will dominate over the lower order Hull--Strominger equations at string scale volumes. No doubt $\ap$--corrections in the form of worldsheet instantons will play an even more important role as well -- it is these that encode geometric invariants of interest in pure mathematics. 

So if we want to apply the formalism of the Hull--Strominger system to examples where $H=\cO(1)$ then we must necessarily include the $\ap$--corrections and $g_s$--corrections to the supersymmetry variations, equations of motion and in turn the moduli space metric. Even if the dilaton is constant, determining how the $\ap$--corrections modify the \K potential beyond the first order result in \citeCOV for heterotic theories remains an open question. These higher order terms will dominate and qualitatively change the behaviour of the metric. For example, such terms should presumably force the metric to remain positive definite even though we have a contribution at order $\ap$ with a negative sign. We know this must be the case as there are strong arguments from \cite{Witten:1985bz,Witten:1986kg} that at least in $\ap$--perturbation theory these geometries define sigma models that flow to unitary conformal field theories with positive definite kinetic terms for the moduli fields, and so, positive definite metric. 

The role of the dilaton is even more mysterious: if there are string solutions on compact manifolds, then they are not described by conventional sigma models. As pointed out in \cite{Witten:2005px}, the $(0,2)$ sigma model for the Hopf surface -- a simple geometric example in which the dilaton is non--trivial at zeroth order in $\ap$ -- is not unitary and so the sigma model does not describe a physical string theory. In fact, the dilaton scalar is not even globally well-defined on the Hopf surface, so it is not clear how it solves the supergravity equations of motion \eqref{EOM} globally. Finding a suitable modification of this worldsheet theory and Hopf surface geometry so that we have a unitary theory flowing to a unitary $(0,2)$ conformal field theory  would be very interesting.  Recently, \cite{Alvarez-Consul:2020hbl} conjectured some of the algebraic structures such a theory should have -- it would be very interesting to try to turn this into a description of a unitary conformal field theory. 

\section{Constructing a family of $\overline{\mathcal{D}}$ operators}
\label{s:Dbops}
We construct a family of $\cDb$--operators on the $Q$ bundle of $(0,p)$--forms that square to zero and $\cDb \ccY_\a^{(1)} = 0$ corresponds to the F-terms; $\cDb^\dag \ccY_\a^{(1)} = 0$ corresponds to the D-terms. 

\subsection{Warm-up: complex manifolds and holomorphic bundles}
As a warm-up we start by setting $\cZ^{(p)}_\a = 0$ and $\d_\a \Th^{(p)} = 0$. The F-term equations \eqref{eq:modulieqnHolGauge}  restrict to
\beq
\begin{split}
\delb_\A (\fD_\a \A) + F_\m  \D_\a{}^\m \= 0~,\qquad \delb \D_{\a}{}^\m \= 0~, \\
\end{split}
\eeq
while the D-term equations are \eqref{eq:adjoint1}--\eqref{eq:adjoint2}.
\beq\label{eq:dtermhym}
\begin{split}
0 & \= \delb_\A^\dag (\fD_\a \A )    ~,\qquad  0 \=\!- \frac{\ap}{4}\tr\big(F^{\nu\mb} \,\fD_\a \A_{\mb}\big)  +\,\nabla^{\mb}\Delta_{\a\mb}{}^\n~. 
\end{split}
\eeq
We define $\Db_1$ to be
$$
\Db_1 \= 
\begin{pmatrix}
 \delb_\A & \cF \\
 0 & \wt\m_p \delb
\end{pmatrix} ~, \qquad \qquad \cF(\D^p_\a) \= \wt \l_p F_\m \D^p_\a{}^\m ~.
$$
with $\wt\l_p, \wt \m_p$ arbitrary coefficients. We see that  $\Db_1^2= 0$
amounts to
 $$
 \delb_\A \wt\l_p F_\m \D_\a^p{}^\m + \wt\l_{p+1} \wt\m_p F_\m \delb \D_\a^p{}^\m~,
 $$
 and so 
 \beq\label{eq:Db1sq}
 -\wt\l_p + \wt\l_{p+1} \wt\m_p \= 0~.
 \eeq
 
We use the appropriate restriction of moduli space metric \eqref{eq:bigmodulimetricspurious}
\beq\notag
\begin{split}
 \langle\ccY^{(p)}_\a , \ccY^{(p)}_\b \rangle &\=  \frac{1}{V}\int_X \left( - \frac{\ap}{4}\tr{ ( \d_\a \A^{(p)} \star \d_{\bb}\A^{(p)}{}^\dag} ) + \D^{(p)}_\a{}^\m\star\D^{(p)}_{\bb}{}^{\nb}\,g_{\m\nb} \right)~.
\end{split}
\eeq
 to compute the adjoint operator
\beq	
\begin{split}
 \langle \Db_1 \ccY_\a^{(p)}, \ccY_\b^{(p+1)} \rangle 
&\= \frac{1}{V} \int_\ccX \Big\{- \frac{\ap}{4} \tr \left( \d_\a \A^{(p)} \star \left( \del^\dag_{\A^\dag} \d_\bb \A^{\dag\,(p+1)} \right) \right)\\
&~~+\left. \D^{(p)}_\a{}^\m \star \left( - \wt\l_p \frac{\ap}{4}  \tr F^{\lb\n}\, \d_\bb \A_\n^\dag{}^{(p+1)}  + \wt\m_p \del^\dag \D^{(p+1)}_{\bb}{}^\lb \right) g_{\m\lb}\right\}\\
&\=  \langle  \ccY_\a^{(p)}, \Db_1^\dag\ccY_\b^{(p+1)} \rangle~,
\end{split}\notag
\eeq
From this we read off
$$
\Db_1^\dag \ccY_\a^{(p+1)} \= 
\begin{pmatrix}
 \delb_\A^\dag \d_\a \A^{(p+1)} \\
  \wt\l_p \frac{\ap}{4}  \tr F^{\m\nb}\, \d_\a \A_\nb{}^{(p+1)}  - \wt\m_p \nabla^\nb \D^{(p+1)}_{\a\,\nb}{}^\m
\end{pmatrix}~.
$$
We check that $(\Db_1^\dag)^2 = 0$. Indeed, the only non--trivial part is 
$$
(\Db_1^\dag)^2 \= (-\wt\l_{p-1} + \wt\l_p \wt\m_{p-1} ) \frac{\ap}{4} \tr F^{\m\tb} \nabla^\lb \d_\a \A_{\tb\lb}^{(p+1)} \= 0~,
$$
and as expected we find exactly \eqref{eq:Db1sq}, the same condition on the coefficients as for $\Db_1^2 = 0$. This is a good consistency check.

We now compare with the F-terms and D-terms. The F-terms amount fo $\Db_1 \ccY_\a^{(1)} = 0$ and so this fixes $\wt\l_1 = 1$. The D-term equation \eqref{eq:dtermhym} corresponds to $\Db_1^\dag \ccY^{(1)}_\a = 0$ and we see
$$
\wt\l_0 = \wt\m_0. 
$$
One simple solution is to pick $\wt\m_p =  \wt \l_p = 1$. We generalise this calculation to include the spin connection and hermitian terms.

\subsection{Hull--Strominger with spurious degrees of freedom}
\label{s:DbsqSpur}
We now do this for the entire Hull--Strominger system with the spurious degrees of freedom. We introduce a $\Db$ operator with $p$-graded coefficients in the same way. Its action in components is
\beq\label{eq:gradedDb}
\begin{split}
 \wh \ccZ_{\a\,\n}^{(p+1)} &\= \m_p \delb \ccZ_{\a\,\n}^{(p)} + 2\ii \ve_p \D_\a^{(p)\,\m}(\del\o)_{\m\n} + \frac{\ap}{2} \l_p \Big( \tr (\d_\a \A^{(p)} F_\n) -  \tr (\d_\a \th^{(p)} R_\n)\Big)~,\\
\wh  \d_\a \A^{(p+1)} &\= \delb_\A \d_\a \A^{(p)} + \wt \l_p F_\m \D_\a^{(p)\,\m}~,\\
\wh  \d_\a \th^{(p+1)} &\= \delb_\th \d_\a \th^{(p)} + \wt \l_p R_\m \D_\a^{(p)\,\m}~,\\
\wh \D_\a^{(p+1)} &\= \wt \m_p \delb \D_\a^{(p)} ~.
\end{split}
\eeq
As a matrix we write this as
\beq
\begin{split}
 \Db &\= 
\begin{pmatrix}
\m_p \delb & \cH_p \\
0 & \Db_2
\end{pmatrix}~,
\end{split}
\eeq
where $\cH_p (\D_\a^{(p)} , \d_\a \A^{(p)} ) \=  2\ii \ve_p \D_\a^{(p)\,\m}(\del\o)_{\m\n} + \frac{\ap}{2} \l_p \left( \tr (\d_\a \A^{(p)} F_\n) - \tr (\d_\a \th^{(p)} R_\n) \right)$ and $\Db_2$ is the natural generalisation of $\Db_1$ taken to act on both $\d \A^{(p)}$ and $\d \th^{(p)}$. 

\beq\label{eq:DbKer}
\Db \begin{pmatrix}
\ccZ_{\a\,\n}^{(p)} \\
\d_\a \th^{(p)} \\
 \d_\a \A^{(p)} \\
 \D_\a^{(p)}
\end{pmatrix} \=  \begin{pmatrix}
 -\delb \ccZ_{\a\,\n}^{(p)} + 2\ii \D_\a^{(p)\,\m}(\del\o)_{\m\n} + \frac{\ap}{2} \Big( \tr (\d_\a \A^{(p)} F_\n) -  \tr (\d_\a \th^{(p)} R_\n)\Big)  \\
\delb_\th( \fD_\a \th)  + \cR(\D_\a)\\
\delb_\A (\fD_\a \A ) + \cF(\D_\a) \\
\delb \D_\a
\end{pmatrix}
~.
\eeq

We now check that $\Db^2 = 0$. From the previous subsection we see that $\Db_2^2 = 0$ amounts to $-\wt\l_p + \wt \l_{p+1} \wt\m_p = 0$. The rest of the calculation amounts to checking 
$$
 \wh \ccZ_{\a\,\n}^{(p+2)} \= \m_{p+1} \delb \wh\ccZ_{\a\,\n}^{(p+1)} + 2\ii \ve_{p+1} \wh \D_\a^{(p+1)\,\m}(\del\o)_{\m\n} + \frac{\ap}{2} \l_{p+1} \Big( \tr \wh \d_\a \A^{(p+1)} F_\n) -  \tr (\wh \d_\a \th^{(p+1)} R_\n)\Big) \= 0~,\\\
$$
We find this is the case if 
\beq\label{eq:dbsq}
\begin{split}
 \l_{p+1} + \l_p \m_{p+1} &\=0~, \quad  \wt\l_p - \wt \l_{p+1} \wt\m_p \= 0~,  \\
 \ve_p \m_{p+1} + \ve_{p+1} \wt \m_p &\= 0 ~,\quad  \m_{p+1} \ve_p + \l_{p+1} \wt\l_p\=0~,
\end{split}
\eeq
together with the Bianchi identity 
$$
2\ii \del\delb \o \=  \frac{\ap}{4} \tr F^2 - \frac{\ap}{4} \tr R^2~.
$$
and we have included the condition for $\Db_2^2 = 0$. Just as for $\Db_1^2 = 0$ in the previous section, this calculation is offshell --- we do not need to use the equations of motion. 

We now compute the adjoint $\Db^\dag$ using the metric \eqref{eq:bigmodulimetricspurious}. 
\beq
\begin{split}
 \langle \Db \ccY_\a, \ccY_\b\rangle &\=  \frac{1}{4V} \int_X \Big\{ \m_p \delb \ccZ_{\a\,\n}^{(p)} + 2\ii \ve_p \D_\a^{(p)\,\m}(\del\o)_{\m\n} + \frac{\ap}{2} \l_p \Big( \tr (\d_\a \A^{(p)} F_\n) -  \tr (\d_\a \th^{(p)} R_\n)\Big) \Big\} \star \ccZb_\bb^\n\\[2pt]
& ~~- \frac{\ap}{4V} \int_X \left( \delb_\A \d_\a \A^{(p)} + \wt \l_p F_\m \D_\a^{(p)\,\m} \right)\star \d_\bb \A^{\dag\,(p+1)} + \frac{\ap}{4V} \int_X \left( \delb_\th \d_\a \th^{(p)} + \wt \l_p R_\m \D_\a^{(p)\,\m} \right)\star \d_\bb \th^{\dag\,(p+1)}\\[2pt]
& ~~+\frac{1}{V}\int \wt\m_p \delb \D_\a{}^\m \star \D_{\bb}^{(p+1)}{}_\m\\[2pt]
&\= \frac{1}{4V} \int_X \ccZ_{\a\,\n}^{(p)} \star \m_p \del^\dag \ccZb_{\bb}^{\,\n}
- \frac{\ap}{4V} \int_X \d_\a \A^{(p)} \star \Big( \del_\A^\dag \d_\bb \A^\dag{}^{(p+1)} + (-1)^{p+1} \frac{\l_p}{2} F_{\n\tb} \ccZb_\bb^{\n\tb} \Big) \\[2pt]
&  \quad +\frac{\ap}{4V} \int_X \d_\a \th^{(p)} \star \Big( \del_\th^\dag \d_\bb \th^\dag{}^{(p+1)} + (-1)^{p+1} \frac{\l_p}{2} R_{\n\tb} \ccZb_\bb^{\n\tb} \Big) \\[2pt]
&\!\!\!+\frac{1}{V} \int_X \D_\a{}^\m\star \left\{ \frac{\ii}{2} \ve_p (-1)^p (\del\o)_{\m\n\tb} \ccZ_\bb^{\n\tb} - \wt \l_p \frac{\ap}{4} \tr F_{\m\tb} \d_\bb \A^\dag{}^\tb + \wt \l_p \frac{\ap}{4} \tr R_{\m\tb} \d_\bb \th^\dag{}^\tb  + \wt\m_p \del^\dag \D_\bb^{(p+1)}{}_\m \right\}~.
\end{split}
\eeq
So we find 
\beq\label{eq:Dbadjaction}
\Db^\dag 
\begin{pmatrix}
 \ccZ_{\a\,\n}^{(p+1)} \\
 \d_\a \th^{(p+1)} \\
 \d_\a A^{(p+1)} \\
 \D^{(p+1)}_\a{}^\m
\end{pmatrix}
\=
\begin{pmatrix}
\m_p \delb^\dag \ccZ^{(p+1)}_{\a\,\n} \\
 \delb_\th^\dag \d_\a \th^{(p+1)} +\half (-1)^{p} \l_p R^{\r\nb} \ccZ^{(p+1)}_{\a\,\r\nb}\\
  \delb_\A^\dag \d_\a \A^{(p+1)} + \half(-1)^p \l_p F^{\r\nb} \ccZ^{(p+1)}_{\a\,\r\nb}\\
  \frac{\ii}{2}(-1)^{p+1} \ve_p(\delb\o)^{\m\r\lb} \ccZ_{\a\,\r\lb} {+} \frac{\ap}{4}\wt\l_p \left( \tr (F^{\m\rb}\, \d_\a \A^{(p+1)}_\rb){-} \tr (R^{\m\rb} \,\d_\a \th^{(p+1)}_\rb)\right) {+}\wt\m_p \delb^\dag \Delta^{(p+1)}_{\a}{}^\n
\end{pmatrix}~.
\eeq
A non--trivial check is that $(\Db^\dag)^2 = 0$ if and only if $\Db^2 =0$. 

We now compare with the F-terms and the D-terms. The F-terms equations \eqref{eq:modulieqnHolGauge}--\eqref{eq:ccZeomgauge} follow if $\Db \ccY_\a^{(1)} = 0$ and $\m_1= -\ve_1 = -\l_1$ and $\wt \l_1 = 1$. The D-term equations are the variations of the HYM and balanced equations, written down in  \eqref{eq:coclosedZ}, \eqref{eq:adjoint1},  \eqref{eq:adjoint2} and \eqref{eq:adjoint3}.\footnote{Its worth noting the last term of \eqref{eq:adjoint1} does not appear as it would come from the inner product of a pair of $(0,2)$--deformations, $\ccZ_\a^{(0,2)}$, which is $\cO(\ap^2)$ and so dropped.} These correspond to $\Db^\dag \ccY^{(1)} = 0$ with $\l_0 = 1$, $\ve_0 = \wt \l_0 = \wt \m_0$. One can check that these satisfy the consistency equations \eqref{eq:dbsq}. Taken together these consistency checks are non--trivial. 

We list two convenient solutions. The first is $\m_p = \wt\l_p = \wt \m_p = 1$ and $\ve_p = \l_p = (-1)^p$. This corresponds to the $\Db$--operator \citeOS in our conventions. A second choice is $\m_p = -1$ and $\wt\l_=\wt\m_p=\l_p=\ve_p=1$, which avoids any explicit $(-1)^p$ factors. In components this second choice is
\beq\label{eq:DbCpt}
\begin{split}
 \wh \ccZ_{\a\,\n}^{(p+1)} &\= -\delb \ccZ_{\a\,\n}^{(p)} + 2\ii \D_\a^{(p)\,\m}(\del\o)_{\m\n} + \frac{\ap}{2} \Big( \tr (\d_\a \A^{(p)} F_\n) -  \tr (\d_\a \th^{(p)} R_\n)\Big)~,\\
\wh  \d_\a \A^{(p+1)} &\= \delb_\A \d_\a \A^{(p)} +  F_\m \D_\a^{(p)\,\m}~,\\
\wh  \d_\a \th^{(p+1)} &\= \delb_\th \d_\a \th^{(p)} +  R_\m \D_\a^{(p)\,\m}~,\\
\wh \D_\a^{(p+1)} &\=  \delb \D_\a^{(p)} ~.
\end{split}
\eeq
So in summary we have found a family of $\Db$ operators and its adjoint via the moduli space metric. The equations of motion correspond to $\Db \ccY^{(1)} = 0$  and $\Db^\dag \ccY^{(1)} = 0$; in other words the first order deformations in holomorphic gauge correspond to the harmonic representatives of $\Db$. 

The non--spurious case follows in an identical manner to this case after making the appropriate substitution for $\d_\a \th^{(p)}$ and redefinition in $\ccZ_\a$ discussed in the main text.

\newpage
\bibliographystyle{utphys}

\providecommand{\href}[2]{#2}\begingroup\raggedright\endgroup

\end{document}